\documentclass[12pt]{article}
\usepackage{graphicx}
\def\mydate{March 10, 2004}
\def\ignore#1{{}}

\let\oldtheequation=\theequation
\def\doteqs#1{\setcounter{equation}{0}
             \def\theequation{{#1}.\oldtheequation}}
\newcounter{sxn}
\def\sx#1{\addtocounter{sxn}{1} \vskip 1.cm  \goodbreak
\noindent{\large\bf\leftline{\thesxn.~~#1}} \nobreak \vskip -.6cm}
\def\sxn#1{\sx{#1} \doteqs{\thesxn}}

\newcounter{axn}

\date{}

\newdimen\mybaselineskip
\renewcommand{\baselinestretch}{1.25}

\newcommand{\beeq}{\begin{equation}}
\newcommand{\eneq}{\end{equation}}
\newcommand{\beqn}{\begin{eqnarray}}
\newcommand{\eeqn}{\end{eqnarray}}

\def\mybig{\displaystyle \strut }

\def\dd{\partial}
\def\la{\raise.16ex\hbox{$\langle$}\lower.16ex\hbox{}  }
\def\ra{\, \raise.16ex\hbox{$\rangle$}\lower.16ex\hbox{} }
\def\go{\rightarrow}

\def\onehalf{ \hbox{${1\over 2}$} }

\def\Tr{{\rm Tr \,}}

\def\eff{{\rm eff}}

\def\sym{{\rm sym}}
\def\phys{{\rm phys}}

\def\BC{{\rm BC}}

\def\hV{{\hat V}}
\def\min{{\rm min}}

\def\Itwo{{\bf 1}_{2\times 2}}

\def\psibar{ \psi \kern-.65em\raise.6em\hbox{$-$} }
\def\psibarl{ \psi \kern-.65em\raise.6em\hbox{$-$} \lower.6em\hbox{} }

\def\myfrac#1#2{{\mybig #1\over \mybig #2}}

\begin{document}
\thispagestyle{empty}

\baselineskip=12pt

{\small \noindent \mydate    \hfill OU-HET 469/2004}

{\small \hfill  hep-ph/0403106}

\baselineskip=40pt plus 1pt minus 1pt

\vskip 2.5cm

\begin{center}

{\Large \bf Dynamical Gauge Symmetry Breaking and}\\
{\Large \bf Mass Generation on the Orbifold $T^2/Z_2$}\\


\vspace{2.5cm}
\baselineskip=20pt plus 1pt minus 1pt

{\bf  Yutaka Hosotani, Shusaku Noda and Kazunori Takenaga}\\
\vspace{.3cm}
{\small \it Department of Physics, Osaka University,
Toyonaka, Osaka 560-0043, Japan}\\
\end{center}

\vskip 3.cm
\baselineskip=20pt plus 1pt minus 1pt

\begin{abstract}
Dynamical gauge symmetry breaking on the orbifold $T^2/Z_2$ is shown to occur
through quantum dynamics of Wilson line phases.  
Different sets of boundary  conditions on  $T^2/Z_2$ can be
related to each other by Wilson line phases, forming  equivalence classes. 
The effective potential for  Wilson line phases is evaluated at the one loop
level in $SU(2)$ gauge  theory.  Depending on the fermion content, the $SU(2)$
symmetry can  be broken either completely or partially to $U(1)$ without
introducing additional Higgs  scalar fields.  When $SU(2)$ is completely
broken, each of three components of the gauge fields may acquire a  distinct
mass. Masses are generated through the combination of $T^2$
twists and dynamics of Wilson line phases.
\end{abstract}
\newpage


\newpage

\sxn{Introduction}
Recently, much attention has been paid on gauge theory on space-time 
with compact extra dimensions.  Gauge theory on an
orbifold 
has been studied extensively in hoping to resolve the long-standing 
problems in grand unified theory (GUT) such as the gauge hierarchy problem,
the doublet-triplet splitting problem, and the
origin of gauge symmetry breaking.\cite{Antoniadis1}-\cite{orbiGUT1}
One intriguing aspect is the  gauge-Higgs unification  in which Higgs bosons 
are regarded as  a part of extra-dimensional components of gauge 
fields.\cite{Manton1}-\cite{Lim1}

Extra dimensions are often compactified on topological 
manifolds. Reflecting the topology of  extra dimensions, dynamical gauge
symmetry breaking occurs through the Hosotani 
mechanism\cite{YH1, YH2} (gauge symmetry
breaking by the Wilson lines). Extra-dimensional 
components of  gauge
fields  (Wilson line phases) become  dynamical degrees of freedom and
cannot be gauged away. They, in general circumstances, develop
nonvanishing  vacuum expectation values \cite{WilsonL1}-\cite{WilsonL2}. 
The extra-dimensional components of  gauge fields 
act as  Higgs bosons at low energies.  Thus  gauge fields and Higgs
particles are  unified by higher  dimensional gauge invariance. One does not
need to introduce extra Higgs fields to break the gauge symmetry.         

To construct realistic GUT, one can choose  extra dimensions to be an
orbifold, which  naturally appears in superstring theory.  By having an
orbifold in extra dimensions, one can easily accommodate chiral fermions in
the four dimensions, and also rich patterns of gauge symmetry breaking. 
In superstring theory,  extra six dimensions must be 
compactified\cite{WilsonL3}-\cite{holonomy}, and therefore
higher dimensional gauge theory might naturally emerge.

Gauge theory on the simplest orbifold $S^1/Z_2$ has been 
studied extensively from various points of view in the 
literature \cite{S1GUT1}-\cite{HHK}.
In this paper, we extend the analysis to  six-dimensional space-time, where
two  of the space coordinates are compactified on 
the orbifold $T^2/Z_2$.\cite{T2GUT1}
In six dimensions there are Weyl fermions which naturally reduce to
four-dimensional Weyl fermions after $Z_2$ orbifolding.
Other orbifolds such as $T^2/Z_3$ and $T^2/Z_4$ have also been
considered to explain the generation structure and 
violation of discrete symmetry.\cite{T2GUT2}  Our
main aim is to study  dynamics of gauge  symmetry breaking and mass
generation on $T^2/Z_2$. We see that 
dynamics of Wilson line phases can reduce or enhance the
symmetry of  boundary conditions.  Such dynamical aspects of gauge
symmetry breaking  have been studied well on $S^1$, $T^n$, and $S^1/Z_2$.  
Effects of supersymmetry breaking and finite masses of matter 
on the dynamics of Wilson line phases have 
been analysed.\cite{Takenaga1, HHHK, Takenaga2}  Dynamics for selecting
boundary conditions are also discussed.\cite{YH4, HHK}
Our analysis given in this paper is expected to provide useful hints for
building realistic unified gauge theory on orbifold to incorporate
the electroweak gauge symmetry breaking within the framework of 
the gauge-Higgs unification.\cite{gaugeHiggs1}-\cite{EWbreaking}

In the next section we classify boundary conditions of fields on the 
orbifold $T^2/Z_2$, and introduce the notion of equivalence classes of  
boundary conditions\cite{YH2, HHHK, YH4, HHK}.  Those equivalence classes 
are connected with the
existence of  Wilson line degrees of freedom.  
$SU(2)$ gauge theory is investigated in detail.   
The effective potential for  Wilson  line phases is evaluated  in 
sections 3 and 4 . In section 5 we examine  gauge symmetry breaking
in the presence of matter fields in various
representations of the gauge group and determine physical symmetry
at low energies.  It is found that depending on matter content, the $SU(2)$
gauge  symmetry is either completely broken or partially 
broken.   It should be emphasized that this makes it plausible to have  
 the electroweak symmetry 
breaking, $SU(2)\times U(1) \rightarrow U(1)_{em}$ 
as a part of the Hosotani mechanism.
In section 6 
we  discuss the masses of  four-dimensional gauge fields, scalar fields and
fermions. Scalar fields, which
are originally the extra-dimensional  components of  gauge fields,  acquire  masses
by radiative corrections.  The final section is devoted to conclusions and
discussion. 


\sxn{Orbifold conditions on $T^2/Z_2$}

We study gauge theory on $M^4\times T^2/Z_2$, where $M^4$ is 
the four-dimensional Minkowski space-time. Let $x^{\mu}$ and $y^I$ be 
coordinates of $M^4$ and $T^2/Z_2$, respectively. The size  
of the two extra dimensions is denoted by $R_I$ $(I=1, 2)$. 
The orbifold $T^2/Z_2$ is given by identifying a 
point $(x^{\mu}, y^I)$ with a point $(x^{\mu}, y^I + 2\pi R_I )$ for 
each $I(=1, 2)$ and further identifying $(x^{\mu}, -y^I)$ 
and $(x^{\mu}, y^I)$. The resultant extra-dimensional space is 
the domain $0\leq y^1 \leq \pi R_1,~0\leq y^2 \leq 2\pi R_2$
with four  
fixed points, $(y^1, y^2)=(0, 0), (\pi R_1, 0), 
(0, \pi R_2), (\pi R_1, \pi R_2)$.

In order for quantum field theory to be defined on space-time 
with compactified spaces,  boundary conditions of 
fields in the compactified dimensions must be specified. In our case we need to
specify  boundary conditions on  $T^2$  
and  for the $Z_2$ orbifolding.  As a general guiding principle  we require
that the  Lagrangian density is singlevalued.  In
gauge theory fields can be twisted up to  gauge degrees of freedom
when they are parallel-transported along noncontractible loops.

\vskip .5cm
\leftline{\bf 2.1 Gauge field}

Let us first consider  boundary conditions for the gauge 
field $A_M(x, y^I)$. The index $M$ runs from $0$ to $5$. 
We define boundary conditions of the gauge potential 
along noncontractible loops on  $T^2$  by
\beqn
T^2: && A_M(x, \vec y + \vec l_a) =
U_a A_M(x, \vec y ) \, U_a^\dagger \quad (a=1,2) \cr
\noalign{\kern 10pt}
&&
\vec l_1 = \pmatrix{2\pi R_1 \cr 0 \cr} ~~,~~
\vec l_2 = \pmatrix{0 \cr 2\pi R_2 \cr} ~~,
\label{BC1}
\eeqn
where $U_I$ $(I=1, 2)$ denote global gauge degrees of freedom 
associated with the original gauge invariance. Gauge potentials at 
$A_M(x,y^1+2\pi R_1, y^2+2\pi R_2)$ is related to 
$A_M(x,y^1, y^2)$ either by a loop translation in the
$y^1$-direction followed by 
a loop translation in the $y^2$-direction, or by a loop translation 
in the  $y^2$-direction followed by 
a loop translation in the $y^1$-direction.  From the consistency it follows that
\begin{equation}
[U_1,~U_2]=0 ~~.
\label{BC2}
\end{equation}

Let us next consider boundary conditions resulting from the $Z_2$ orbifolding. 
To simplify expressions, we denote four fixed points on $T^2/Z_2$ by 
$\vec z_i$ ($i=0, 1, 2, 3$);
\beeq
\vec z_0 = \pmatrix{0 \cr 0 \cr} ~~,~~
\vec z_1 = \pmatrix{\pi R_1 \cr 0 \cr} ~~,~~
\vec z_2 = \pmatrix{0 \cr \pi R_2 \cr} ~~,~~
\vec z_3 = \pmatrix{\pi R_1 \cr \pi R_2 \cr} ~~.
\label{FixedP}
\eneq
Boundary conditions are specified by 
unitary parity matrices $P_i$ $ (i=0, 1, 2, 3)$  at the
fixed  points;
\beeq
Z_2: \quad
\pmatrix{A_\mu \cr A_{y^I} \cr} (x, \vec z_i - \vec y) =
P_i \pmatrix{A_\mu \cr - A_{y^I} \cr} (x, \vec z_i + \vec y) \, 
   P_i^\dagger \quad (i=0,1,2,3) ~~.
\label{BC3}
\eneq
$A_{y^I}(I=1, 2)$ must have an opposite sign relative to $A_{\mu}$ under 
these transformations in order to preserve the gauge invariance. 
The repeated $Z_2$ parity operation brings a field configuration back 
to the original one, so that $P_i^2=1$ $(i=0, 1, 2, 3)$ and 
hence, $P_i^{\dagger}=P_i$. 

At this stage, we observe that not all the boundary conditions are 
independent. The transformation $\pi R_1 -y^1 \rightarrow \pi R_1+ y^1 $ must 
be the same as $\pi R_1 -y^1 \rightarrow -\pi R_1 +y^1 
\rightarrow y^1 + \pi R_1$, from which it follows that $U_1=P_1P_0$.  A similar
relation holds for $U_2$.  We have
\beeq
U_a=P_a P_0  \quad (a=1,2) ~.
\label{BC4}
\eneq 
Finally, as the transformation $(\pi R_1 -y^1, \pi R_2 -y^2)
\rightarrow (\pi R_1 +y^1, \pi R_2 +y^2)$ is the same as a transformation 
$(\pi R_1 -y^1, \pi R_2 -y^2)   \rightarrow 
(-\pi R_1 +y^1, -\pi R_2 + y^2) \rightarrow 
(\pi R_1 + y^1, -\pi R_2 +y^2)  \rightarrow 
(\pi R_1 +y^1, \pi R_2 +y^2)$, the relation $U_2U_1P_0=P_3$ must hold.
Taking account of Eqs. (2.4), (2.5) and (2.6),  the parity matrix $P_3$ 
can be written as
\begin{equation}
P_3=P_2P_0P_1=P_1P_0P_2 ~~.
\label{BC5}
\end{equation}
The boundary conditions for gauge fields are specified with $P_i ~ (i=0, 1, 2)$
satisfying $P_i=P_i^\dagger = P_i^{-1}$ and $P_i P_j P_k = P_k P_j P_i$.
    
Discussions can be generalized to the case 
of $T^n/Z_2$. The orbifold $T^n/Z_2$ is defined by identifications,
\begin{eqnarray} 
T^n : &&  \vec{y}+\vec{l}_j \sim \vec{y}~~(j=1, 2, \cdots, n),\\
Z_2 : && -\vec{y}\sim \vec{y},
\end{eqnarray}
where $\vec{y}$ is an $n$-dimensional vector on the $n$-torus 
and $\vec{l}_j \equiv (0, \cdots, 0, 2\pi R_j, 0, \cdots, 0)^T~~(j=1, \cdots, n).$
The fixed point satisfies the 
relation $\vec{y}=-\vec{y}+\sum_j m_j\vec{l}_j$ ($m_j=$ an integer).
In the fundamental domain of $T^n$, they are given by
$\vec{y}={1\over 2}\sum_j m_j \vec{l}_j$ where $m_j =$ 0 or 1.
In the $T^2/Z_2$ case, there are four fixed points corresponding to 
$(m_1, m_2)=(0, 0), (0, 1), (1, 0), (1, 1)$. 
At each fixed point the parity matrix is  defined. Repeating 
the same discussion given above, $n+1$ matrices, for 
example, $P_0, P_1, \cdots, P_n$ are independent. 
The consistency of the $Z_2$ orbifolding 
and the $T^n$ boundary condition defined by $U_j$ $(j=1, \cdots, n)$
satisfying $[U_j,~U_k]=0 (j\neq k)$ 
yields the relation $U_j=P_j P_0$.  The relation $P_i P_j P_k = P_k P_j P_i$ also
holds.

\vskip .5cm
\leftline{\bf 2.2 Matter fields}

As for matter fields, it is convenient to first specify  $Z_2$ 
boundary condition and then derive  $T^2$ conditions. Let us consider a
scalar field $H(x, \vec y)$ which satisfies
\beeq
Z_2: \quad 
H(x, \vec z_j - \vec y)=\eta_j^s \, T[P_j] \, H(x, \vec z_j + \vec y) 
\quad (j=0,1,2,3)~~.
\label{BCs1}
\eneq
Here $T[P_j]$ stands for an appropriate representation matrix under 
the gauge group associated with $P_j$. If $H$ belongs to the fundamental or adjoint 
representation, $T[P_j] \, H = P_j H$ or $P_j H P_j^{\dagger}$, respectively. 
$\eta_j^s$ is a sign  factor taking a value $+1$ or $-1$. 
Boundary conditions  for the $T^2$ direction are  given by 
\beeq
T^2: \quad 
H(x, \vec y + \vec l_a) 
= \eta_0^s\eta_a^s \, T[U_a] \, H(x, \vec y) \quad (a=1,2) ~~,
\label{BCs2}
\eneq
where $U_a$ is given by (\ref{BC4}).
Not all of the sign factors are  
independent; $\eta_3^s=\eta_0^s\eta_1^s\eta_2^s$.
 
Next we consider a Dirac fermion $\psi(x, y^I)$ in six 
dimensions. 
The gauge invariance of the kinetic term of the fermion Lagrangian demands that
\beqn
Z_2 : &&
\psi(x, \vec z_j - \vec y) = \eta_j^f \, T[P_j] \, (i\Gamma^4\Gamma^5)
\psi(x, \vec z_j + \vec y) \quad (j=0,1, 2,3) ~~, \cr
\noalign{\kern 5pt}
T^2 : &&
\psi(x, \vec y + \vec l_a) 
  = \eta_0^f \eta_a^f \, T[U_a] \psi(x, \vec y) \quad (a=1,2) ~~.
\label{BCf1}
\eeqn
The sign factors $\eta_j^f=\pm 1$ satisfy $\eta_3^f=\eta_0^f\eta_1^f\eta_2^f$.
$\Gamma^4$ and $\Gamma^5$ are the fourth and fifth components of   
six-dimensional $8\times 8$ Dirac's gamma matrices, respectively. 

It is instructive to present the explicit form of gamma matrices. 
We employ the following representation;
\begin{equation}
\Gamma_{\mu}=\gamma_{\mu}\otimes {\bf 1}_{2\times 2} ~~,\quad 
\Gamma_4 = \gamma_5 \otimes i\sigma_1 ~~,\quad
\Gamma_5 = \gamma_5 \otimes i\sigma_2 ~~,
\label{gamma1}
\end{equation}  
where $\gamma_{\mu}$ is the four-dimensional gamma matrix 
and $\gamma_5\equiv i\gamma_0\gamma_1\gamma_2\gamma_3$ 
with $(\gamma_5)^2={\bf 1}_{4\times 4}$. In this representation 
we have 
\begin{equation}
i\Gamma^4 \Gamma^5 = {\bf 1}_{4\times 4}\otimes \sigma_3 ~~.
\label{gamma2}
\end{equation}
One can define six-dimensional chirality  similar to the 
chirality in four dimensions. It is given by the eigenvalues 
of $\Gamma^7$ defined by
\begin{equation}
\Gamma^7\equiv \Gamma^0\Gamma^1 \cdots \Gamma^5 
(=\gamma_5 \otimes\sigma_3) ~~. 
\label{gamma3}
\end{equation}     
Then, we obtain that
\begin{equation}
\Gamma^7\psi_{\pm} =\pm\psi_{\pm} ~~, 
\quad\mbox{where}\quad \psi_{\pm} \equiv
{1\over 2}(1\pm \Gamma^7)\psi ~~.
\label{chiral1}
\end{equation}
If we write 
\begin{equation}
\psi_{-} =\pmatrix{U_L \cr D_R} ~~,~~ 
\psi_+ =\pmatrix{U_R \cr D_L} ~~,
\label{updown1}
\end{equation}
$\gamma_5 U_L=-U_L, \gamma_5 D_L=-D_L, \gamma_5 U_R =U_R, \gamma_5 D_R =D_R$. 
In terms of four-dimensional Dirac spinors the boundary conditions
(\ref{BCf1})  are recast as
\beqn
Z_2: &&
U_{L,R}  (x, \vec z_j - \vec y)
 = + \eta_j^f \, T[P_j] \, U_{L,R} (x, \vec z_j + \vec y) ~~, \cr
&& D_{L,R}  (x, \vec z_j - \vec y) 
= - \eta_j^f \, T[P_j] \, D_{L,R}  (x, \vec z_j + \vec y) ~~,
\label{BCf2}
\eeqn

The sign factors $\{ \eta_j^s \}$, $\{ \eta_j^f \}$ are additional 
parameters  specifying  boundary conditions. They play an important role in
dynamical gauge symmetry breaking.

\vskip .5cm
\leftline{\bf 2.3 Equivalence classes and symmetry of boundary conditions}

The gauge symmetry is apparently broken by nontrivial parity 
matrices $P_j$ ($j=0, 1, 2)$ specifying boundary conditions of 
the $Z_2$ orbifolding.  Yet physical symmetry of the theory is not,
in general, the same as the symmetry of boundary conditions, once quantum
corrections are incorporated.

To elucidate this fact, we first show that different sets of boundary 
conditions can be related to each other by `large' gauge transformations.
Under a gauge transformation
\beeq
A_M'
=\Omega \bigg( A_M -{i\over g}\partial_M\bigg) \Omega^{\dagger}
\label{gaugeT1}
\end{equation}
$A_M'$ obeys a new set of boundary conditions $\{ P_j' , U_a' \}$ where
\beqn
&&\hskip -1cm
P_j' =\Omega(x, \vec z_j -\vec y) \, P_j 
   \,\Omega(x, \vec z_j +\vec y)^\dagger ~, \cr
\noalign{\kern 5pt}
&&\hskip -1cm
U_a' = \Omega(x, \vec y + \vec l_a)\, U_a\, \Omega(x, \vec y)^\dagger ~, \cr
\noalign{\kern 5pt}
&&\hskip 0. cm
\hbox{provided ~} \dd_M P_j' = \dd_M U_a' = 0 ~~.
\label{newBC1}
\eeqn 
The relation $U_a' = P_a' P_0'$ follows from (\ref{BC4}) 
and (\ref{newBC1}).  We stress that the set $\{ P_j' \}$ can be different
from the set $\{ P_j \}$.  When the relations in (\ref{newBC1}) are
satisfied, we write
\beeq
\{ P_j' \} \sim \{ P_j \} ~~.
\label{relation1}
\eneq
This relation is transitive, and therefore is an equivalence
relation.  Sets of boundary conditions form equivalence classes of boundary 
conditions with respect to the equivalence 
relation (\ref{relation1}). \cite{YH2, HHHK,  HHK}

The residual gauge invariance of the 
boundary conditions is given by  gauge transformations that 
preserve the original boundary conditions;
\beqn
&&\hskip -1cm
P_j =\Omega(x, \vec z_j -\vec y) \, P_j 
   \,\Omega(x, \vec z_j +\vec y)^\dagger ~, \cr
\noalign{\kern 5pt}
&&\hskip -1cm
U_a = \Omega(x, \vec y + \vec l_a)\, U_a\, \Omega(x, \vec y)^\dagger ~.
\label{residual1}
\eeqn 
 As shown in \cite{HHHK}, those
residual gauge transformations extend over the entire group space
even for nontrivial $\{ P_j \}$.  All the Kaluza-Klein modes nontrivially
mix under those gauge transformations.

The gauge symmetry realized at low energies is given 
by $y^I$-independent $\Omega$ satisfying
\begin{equation}
[P_j,~\Omega(x)]=0  \quad (j=0, 1, 2) ~.
\label{residual2}
\end{equation} 
We observe that the symmetry is generated by  generators of 
the gauge group which commute with $P_j$. This is 
called the symmetry of boundary conditions at low energies. 

The  gauge symmetry at low energies 
can also be understood in terms of  group generators associated with  zero
modes of the gauge fields $A_{\mu} =  A_\mu^a T^a$.  
Let us define
\beqn
&&\hskip -1cm
{\cal H}_{\rm BC} = \Big\{ ~ T^a ~;~ [T^a,~P_j] =0 ~~(j=0,1,2) \Big\} ~~,\cr
\noalign{\kern 5pt}
&&\hskip -1cm
{\overline{\cal H}}_{\rm BC} = \Big\{ ~ T^b ~;~ \{ T^b ,~P_j\}
     =0 ~~(j=0,1,2) \Big\} ~~.
\label{set1}
\eeqn
From the boundary condition (\ref{BC3}),  it follows that zero modes
($y^I$-independent modes) of $A_\mu$ and $A_{y^I}$ can be written as
\beqn
&&\hskip -1cm
A_\mu(x) = \sum_{T^a \in {\cal H}_{\BC}}  A_\mu^a (x)\, T^a ~~, 
\label{zeromode1} \\
\noalign{\kern 5pt}
&&\hskip -1cm
A_{y^I} (x) = \sum_{T^b \in {\overline{\cal H}}_{\BC}} 
    A_{y^I}^b (x)  \, T^b ~~.
\label{zeromode2}
\eeqn
The residual gauge symmetry at low energies $H_\BC$ is spanned by those
generators belonging to ${\cal H}_\BC$.  

The zero modes $A_{y^I}$ in (\ref{zeromode2}), or
particularly their $x$-independent parts, define Wilson line phases and
play a critical role in dynamical rearrangement of gauge symmetry at
the quantum level, which we elaborate in the following subsection.

\vskip .5cm
\leftline{\bf 2.4 Wilson line phases and physical symmetry}

So far we have discussed the symmetry of the boundary condition 
$\{ P_j \}$ at the tree level.  This is not necessarily the same as the 
physical symmetry of the theory.  Once quantum corrections are taken into
account, the boundary condition effectively changes as a result of $A_{y^I}$
in (\ref{zeromode2}) developing nonvanishing expectation values.
The number of zero modes of four-dimensional gauge fields $A_{\mu}^a$ 
in the new vacuum also changes.  Rearrangement of gauge symmetry takes place.
This is called the Hosotani mechanism\cite{YH1, YH2}.

Constant modes of $A_{y^I}$ satisfying $[A_{y^1} , A_{y^2} ]=0$ give
vanishing field strengths, but become physical degrees of freedom that
cannot be gauged away within the given boundary conditions.  Indeed
the path-ordered integral along a noncontractible loop starting at $(x,y)$
\begin{equation}
W_{I}(x,y) ={\cal P} \exp \left( 
  ig \oint dy^I\,   A_{y^I}  \right) ~,
\quad  (I: ~\mbox{not~summed}),
\end{equation}
transforms, under a gauge transformation, as 
$W_{I}(x,\vec y) \go \Omega(x,\vec y) W_{I}(x,\vec y) 
  \Omega(x, \vec y + \vec l_I)^\dagger$.  Using (\ref{residual1}), one 
finds that
\begin{equation}
W_I (x,\vec y) U_I \go \Omega (x,\vec y) \,  W_I (x,\vec y) U_I \,
   \Omega^{\dagger}(x,\vec y)  ~~.
\label{wilson1}
\end{equation}
The eigenvalues of $W_I U_I $ are invariant under gauge 
transformations preserving the boundary conditions.  The
phases of the  eigenvalues, called Wilson line phases, cannot be gauge away. 
They are non-Abelian analogues of  Aharonov-Bohm phases.  

These Wilson line phases parametrize degenerate classical vacua. 
At the quantum
level the effective potential for Wilson line phases becomes nontrivial.
When the effective potential is minimized at nonvanishing Wilson line 
phases, the physical symmetry of the theory changes from the symmetry of
boundary conditions.

The effect of nonvanishing vacuum expectation values of Wilson line
phases can be understood as an effective change in boundary conditions. 
As explained in the previous subsection, there are large gauge
transformations which change boundary conditions.  The existence of such
gauge transformations is in one-to-one correspondence with the existence of
physical degrees of freedom of Wilson line phases in a  given theory. 

Suppose that the effective potential is minimized at nonvanishing 
$\la A_{y^I} \ra \not= 0$, $[\la A_{y^1} \ra, \la A_{y^2} \ra ]$=0.  
Perform a large gauge transformation
\beeq
\Omega(\vec y) = \exp \bigg\{ - ig \Big( \la A_{y^1} \ra y^1 +
   \la A_{y^2} \ra y^2 \Big) \bigg\} ~~.
\label{largeGT1}
\eneq
Then the new gauge potentials satisfy $\la A_{y^I}' \ra = 0$.  
Simultaneously the boundary conditions change as in (\ref{newBC1});
\beqn
P_j' &=& \Omega(\vec z_j -\vec y) P_j \Omega(\vec z_j + \vec y)^\dagger \cr
&=& P_j \Omega(-\vec z_j +\vec y) \Omega(\vec z_j + \vec y)^\dagger \cr
&=& P_j \Omega(-2\vec z_j) 
\hskip 1cm \equiv P_j^\sym ~~~.
\label{physBC1}
\eeqn
In the second equality we made use of the relation
$\{ \la A_{y^I} \ra , P_j \} = 0$.  Since  $\la A_{y^I}' \ra = 0$,
the physical symmetry of the theory $H_\phys$ is generated by generators
belonging to
\beeq
{\cal H}_\phys = \Big\{ ~ T^a ~;~ [T^a,~P_j^\sym ] =0 ~~(j=0,1,2) \Big\} ~~.
\label{set2}
\eneq
The physical symmetry ${H}_\phys$ can be either larger or smaller
than ${H}_\BC$.

As emphasized in Refs.\ \cite{HHHK} and \cite{HHK}, the physical symmetry 
$H_\phys$ is the same in all theories belonging to the same equivalence 
class of boundary conditions.  Dynamics of Wilson line phases 
guarantee it.

\sxn{Orbifold conditions and mode expansions in $SU(2)$ theory}

Let us examine $SU(2)$ gauge theory for which complete classification
of orbifold boundary conditions can be easily achieved.
Boundary condition matrices  $P_j$ ($j=0,1,2$) must satisfy
$P_j = P_j^\dagger = P_j^{-1}$ and $P_1 P_0 P_2 = P_2 P_0 P_1$.
Complete classification of boundary conditions in $SU(N)$ gauge theory on 
the orbifold $S^1/Z_2$ has been given in ref.\ \cite{HHK}. 

\vskip .3cm

\leftline{\bf 3.1 Orbifold conditions}

To classify  boundary conditions $\{ P_j \}$, we first diagonalize $P_0$,
utilizing global $SU(2)$ invariance.  Up to a sign factor, 
$P_0 = \Itwo$ or $\tau^3$.  If $P_0 = \Itwo$, $P_1$ can be diagonalized,
and therefore $P_1 = \Itwo$ or $\tau^3$.  In the case $P_0=P_1=\Itwo$, $P_2$
is diagonalized as well.  Even in the case $P_0=\Itwo$, $P_1=\tau^3$, 
$P_2$ must be diagonal to satisfy $P_1 P_2 = P_2 P_1$.  In other words,
if one of $P_j$'s is $\Itwo$, all $P_j$'s are diagonal up to a global 
$SU(2)$ transformation.  

In the case $P_0= \tau^3$ and $P_1,P_2 \not= \pm \Itwo$, we recall that the
most general form of
$P (\not= \pm \Itwo)$ satisfying $P=P^\dagger=P^{-1}$ is given by 
$P = \tau^3 e^{i(\alpha_1 \tau^1 + \alpha_2 \tau^2)}$. 
Given $P_0= \tau^3$, there still remains  $U(1)$ invariance.  Utilizing the
global $U(1)$ invariance, one can bring $P_1$ into the form 
$P_1= \tau^3 e^{i\pi a \tau^2}$.   Then, to satisfy 
$P_1 \tau^3 P_2= P_2 \tau^3 P_1$, $P_2$ must be 
$P_2= \tau^3 e^{i\pi b \tau^2}$.

To summarize,  boundary conditions $\{ P_0, P_1, P_2 \}$ are classified as
\beqn
&{\rm (i)}& P_0  =  \Itwo ~,~ P_1, P_2  =  \Itwo ~{\rm or}~ \pm \tau^3  ~~~, 
\label{BC6s}  \\
\noalign{\kern 10pt}
&{\rm (ii)}& (P_0 , P_1, P_2) 
  = (\tau^3 , \tau^3 e^{i\pi a \tau^2}, \tau^3 e^{i\pi b \tau^2}) \cr
\noalign{\kern 5pt}
&&\hskip .2cm 
P_3 = \tau^3 e^{i\pi (a+b) \tau^2} ~~, \cr
\noalign{\kern 5pt}
&&\hskip .2cm 
U_1 = e^{-i\pi a \tau^2} 
    = \pmatrix{\cos \pi a & -\sin \pi a \cr
															\sin \pi a & \cos \pi a }  ~~, ~~
U_2 =  \pmatrix{\cos \pi b & -\sin \pi b \cr
															\sin \pi b & \cos \pi b } ~~.
\label{BC6d}
\eeqn
The boundary condition (\ref{BC6d}) is periodic in real parameters $a, b$
with a period 2.
The symmetry of the boundary condition (\ref{BC6s}) is either $SU(2)$ or
$U(1)$. The symmetry of the boundary condition (\ref{BC6d}) is $U(1)$ if
both $a$ and $b$ are integers, and none otherwise.

\vskip .5cm
\leftline{\bf 3.2 Wilson line phases}

There is no degree of freedom of a Wilson line phase with the boundary
condition (\ref{BC6s}).  In the case of the boundary condition (\ref{BC6d})
with general values of $a$ and $b$,  there is no zero 
mode associated with $A_{\mu}^a$, but there are zero modes for $A_{y^I}$
and may develop expectation values;
\beeq
\la A_{y^1}\ra ={\alpha\over{2R_1 g}}\tau^2  ~~,~~
\la A_{y^2}  \ra ={\beta\over{2R_2 g}}\tau^2 ~~.
\label{wilson2}
\eneq 
The expectation values $\alpha$ and $\beta$ are dynamically determined
such that the effective potential is minimized.  They are  related to the
Wilson line phases by
\beqn
&&\hskip -1cm
W_1 U_1= \la e^{ig \oint dy^1 \, A_{y^1} } \ra \cdot e^{-i\pi a \tau^2}
=e^{i \pi(\alpha-a) \tau^2} ~~,
\cr 
\noalign{\kern 5pt}
&&\hskip -1cm
W_2 U_2= \la e^{ig \oint dy^2 \, A_{y^2} } \ra \cdot e^{-i\pi b \tau^2}
=e^{i \pi(\beta - b) \tau^2} ~~.
\label{wilson3}
\eeqn

\vskip .5cm
\leftline{\bf3.3  Equivalence classes of boundary conditions}

Consider the boundary condition in (\ref{BC6d}).  We perform 
a large gauge transformation with
\begin{equation}
\Omega(c_1, c_2)=
\exp \bigg\{ i\Big( {c_1 \over{2R_1}}y^1 
   + {c_2 \over{2R_2}}y^2 \Big) \tau^2 \bigg\} ~~.
\label{largeGT2}
\end{equation}
Then the boundary condition matrices change to 
\begin{equation}
(P_0^{\prime}, P_1^{\prime}, P_2^{\prime})
=(\tau^3, \tau^3 e^{i\pi(a-c_1)\tau^2} , 
\tau^3 e^{i\pi (b-c_2)\tau^2})  ~~. 
\end{equation}   
In other words, all sets $(P_0, P_1, P_2)$ of  boundary conditions in
(\ref{BC6d}) are in one equivalence class of boundary conditions.
Each set of the boundary conditions in (\ref{BC6s})  forms a distinct
equivalence class.

Under (\ref{largeGT2}), the zero modes of $A_{y^I}$ in (\ref{wilson2})
are transformed as $(\alpha, \beta) \go (\alpha -c_1, \beta-c_2)$.
It is recognized that the combination  $(\alpha -a, \beta-b)$ is invariant 
under (\ref{largeGT2}).

Now suppose that the expectation values $\alpha, \beta$ in (\ref{wilson2}) 
take nontrivial values.  With a gauge transformation $\Omega(\alpha, \beta)$,
the background field $\langle A_{y^I}\rangle$ can be removed, and in 
the new gauge we have $\langle A_{y^I}^{\prime}\rangle =0$.
The new boundary conditions are
$(P_0^\sym, P_1^\sym, P_2^\sym)
=(\tau^3, \tau^3 e^{i\pi(a-\alpha)\tau^2}, 
\tau^3 e^{i\pi (b-\beta)\tau^2})$.
The physical symmetry $H_\phys$ is generated by the generators of the $SU(2)$
commuting  with $P_{i}^\sym$ $(i=0, 1, 2)$.

Physical content of the theory at the quantum level is the same in a given 
equivalence class.  In particular, it does not depend on the parameters
$(a,b)$ in (\ref{BC6d}).  Gauge invariance implies that the effective
potential for the Wilson line phases is a function of gauge invariant
$\alpha-a$ and $\beta-b$.  Depending on the content of matter fields,
the effective potential can take the minimum value at nontrivial 
$(\alpha-a ,\beta-b)$ as we will see below.

\vskip .5cm

\leftline{\bf 3.4 Mode expansions}

Given the orbifold boundary conditions, each field is expanded in
eigenmodes.  On $T^2/Z_2$ there are two types of mode expansions,
$Z_2$ singlets and $Z_2$ doublets.

A $Z_2$ singlet field $\phi(x,y)$ obeys 
\beeq
\phi (x, \vec z_j - \vec y) = P_j \phi (x, \vec z_j + \vec y)
~~~,~~~ P_j = + ~{\rm or}~ - ~~.
\label{singlet1}
\eneq
Each singlet field is specified with $(P_0,P_1,P_2)$.  Mode expansions are
\beqn
&&\hskip -1 cm
\phi(x,\vec y) = \cr
\noalign{\kern 10pt}
&&\hskip -1 cm
{1\over \sqrt{2\pi^2 R_1 R_2}}  \phi_{00} (x)  \pmatrix{1 \cr 0}
+ {1\over \sqrt{\pi^2 R_1 R_2}}  \sum_{(n,m) \in K_+}
\phi_{nm} (x)  \pmatrix{\cos \cr \sin }
\left({n y^1\over R_1}+{m y^2\over R_2}\right) 
\cr 
\noalign{\kern 5pt}
&&\hskip 7.0 cm
\hbox{for~} (P_0,P_1,P_2)=
 \left\{ \matrix{(+,+,+)\cr  (-,-,-)\cr} \right. ~,
\label{mode1} \\
\noalign{\kern 10pt}
&&\hskip -1cm
{1\over \sqrt{\pi^2 R_1 R_2}} 
\sum_{n=-\infty}^\infty \sum_{m=0}^\infty
\phi_{nm} (x)  \pmatrix{\cos \cr \sin }
\left( {n y^1\over R_1}
   +{(m+\onehalf) y^2\over R_2}\right)  \cr
\noalign{\kern 5pt}
&&\hskip 7.0 cm
\hbox{for~} (P_0,P_1,P_2)=
 \left\{ \matrix{(+,+,-)\cr  (-,-,+)\cr} \right. ~,
\label{mode2} \\
\noalign{\kern 10pt}
&&\hskip -1 cm
{1\over \sqrt{\pi^2 R_1 R_2}} 
\sum_{n=0}^\infty \sum_{m=-\infty}^\infty
\phi_{nm} (x)  \pmatrix{\cos \cr \sin } 
\left( {(n+\onehalf) y^1\over R_1} +{m y^2\over R_2}\right) \cr
\noalign{\kern 5pt}
&&\hskip 7.0 cm
\hbox{for~} (P_0,P_1,P_2)=
\left\{ \matrix{(+,-,+)\cr  (-,+,-)\cr} \right. ~,
\label{mode3} \\
\noalign{\kern 10pt}
&&\hskip -1cm
{1\over \sqrt{\pi^2 R_1 R_2}} 
\sum_{n=-\infty}^\infty \sum_{m=0}^\infty
\phi_{nm} (x)  
 \pmatrix{\cos \cr \sin } 
\left( {(n+\onehalf) y^1\over R_1}
   +{(m+\onehalf) y^2\over R_2}\right) \cr
\noalign{\kern 5pt}
&&\hskip 7.0 cm
\hbox{for~} (P_0,P_1,P_2)=
 \left\{ \matrix{(+,-,-)\cr  (-,+,+)\cr} \right. ~.
\label{mode4} 
\eeqn
In (\ref{mode1})
\beeq
\sum_{(n,m) \in K_+} B_{n,m} = \sum_{n=1}^\infty B_{n,0} 
 + \sum_{n=-\infty}^\infty \sum_{m=1}^\infty B_{n,m} ~~.
\label{half-sum}
\eneq
Zero modes exist only for $(P_0,P_1,P_2) = (+,+,+)$.

A $Z_2$ doublet field $\phi= \mybig \pmatrix{\phi_1\cr \phi_2}$ appears when
the boundary condition of the type (\ref{BC6d}) is considered. It obeys
\beqn
&&\hskip -1cm
\phi (x, -y^1, -y^2)
 = \pmatrix{1&\cr &-1} \phi (x, y^1, y^2) ~~, \cr
\noalign{\kern 10pt}
&&\hskip -1cm
\phi(x, y^1+2\pi R_1, y^2)
 =\pmatrix{ \cos \pi a & -\sin \pi a\cr \sin \pi a & \cos \pi a} 
\phi(x, y^1, y^2) ~~, \cr
\noalign{\kern 10pt}
&&\hskip -1cm
\phi(x, y^1, y^2 +2\pi R_2)
 =\pmatrix{ \cos \pi b & -\sin \pi b\cr \sin \pi b & \cos \pi b} 
\phi(x, y^1, y^2) ~~.
\label{doublet1}
\eeqn
Its mode expansion is given by
\beeq
\pmatrix{\phi_1 \cr \phi_2} (x, \vec y) = 
{1\over \sqrt{2 \pi^2 R_1 R_2}} 
\sum_{n=-\infty}^\infty \sum_{m=-\infty}^\infty  \phi_{nm} (x) 
\pmatrix{\cos \cr \sin }
\Bigg[ \myfrac{(n+ \onehalf a) y^1}{R_1}
            +\myfrac{(m+\onehalf b) y^2}{ R_2}\Bigg]  ~.
\label{modeD1}
\eneq
$Z_2$ doublets appear when the Scherk-Schwarz SUSY 
breaking \cite{SS, Takenaga1} is implemented
in SUSY theories as well.  We see below that twists specified with $(a,b)$ 
play an important role to give fermions  nonvanishing  masses in 
four dimensions.

\sxn{Effective potential in $SU(2)$ gauge theory}

In order to study physical symmetry of the theory, one must take into 
account quantum corrections. To this end one needs to 
evaluate the effective potential for Wilson line phases.
Wilson line phases  are
related to  zero modes of the component gauge fields in extra dimensions.
 
We study patterns of gauge symmetry breaking in
 $SU(2)$ gauge theory on $M^4\times T^2/Z_2$ in order 
to get insight into the electroweak gauge symmetry breaking and 
the gauge-Higgs unification in a more realistic framework. 
We believe that the 
analysis here provides us many useful and important hints.

Dynamical gauge symmetry breaking or enhancement can take place when the 
boundary condition in (\ref{BC6d}) is adopted.  As explained in the 
previous section, all the boundary conditions in (\ref{BC6d})  are in 
one equivalence class, and therefore the same physical results are obtained
independently of the values of the parameters $(a,b)$, provided that dynamics
of Wilson line phases are taken into account. Hence we adopt, without loss of
generality, 
\begin{equation}
(P_0, P_1, P_2)=(\tau^3, \tau^3, \tau^3) ~~,
\label{BC7}
\end{equation}     
which implies that $P_3=\tau^3$ and $U_1=U_2=\Itwo$.
With this boundary condition, zero modes for $A_{\mu}^a$ 
and $A_{y^I}^a (I=1, 2)$ exist for $A_{\mu}^{a=3}$ 
and $A_{y^I}^{a=1, 2}$ $(I=1, 2)$, respectively. The symmetry 
of the theory at the tree level is $U(1)$. 

We employ the standard background filed method to evaluate the 
effective potential for Wilson line phases. 
Zero modes for $A_{y^I}^{a=1, 2}$ are parameterized as
\begin{eqnarray}
 A_{y^1} &=& {1\over 2gR_1}(\alpha_1\tau^1 + \alpha_2\tau^2)
\equiv{\alpha\over {2gR_1}} \pmatrix{0 & e^{-i\theta}\cr
                e^{i\theta} & 0\cr} ~, \cr
 A_{y^2} &=& {1\over 2gR_2}(\beta_1\tau^1 + \beta_2\tau^2)
\equiv{\beta\over {2gR_2}} \pmatrix{0 & e^{-i\tilde\theta}\cr
                e^{i\tilde\theta} & 0\cr} ~.
\label{wilson4}
\end{eqnarray}
Here one should note that, contrary to the case $S^1/Z_2$, there are two
directions of the compactified dimensions, so that 
the tree-level potential is induced for the background given above,
\begin{equation}
V_{tree}=-{g^2\over 2}{\rm tr}
[ A_{y^1}  , ~   A_{y^2}  ]^2
={1\over {4g^2(R_1R_2)^2}}(\alpha_1\beta_2-\alpha_2\beta_1)^2.
\end{equation}    
 The vanishing tree-level potential is achieved when
\begin{equation}
\alpha_1\beta_2-\alpha_2\beta_1=0  ~~,
\label{wilson5}
\end{equation}  
which implies the vanishing field 
strength $\langle F_{y^1y^2}\rangle =0$. Once we restrict ourselves to 
the case (\ref{wilson5}), the
parameterization of background fields is further simplified. The
relation (\ref{wilson5}) means $\theta=\tilde\theta$ and by using the $U(1)$
gauge degrees  of freedom, one can take $\theta=\tilde\theta= 0$.
To summarize, we take, as background fields, 
\begin{equation}
 A_{y^1}  = {1\over 2gR_1}\alpha\tau^1 ~~,\quad
 A_{y^2}  = {1\over 2gR_2}\beta\tau^1 ~~.
\label{wilson6}
\end{equation}
The effective potential for $(\alpha, \beta)$ is obtained by  integrating
 quantum fluctuations of every field.

\vskip .5cm
\leftline{\bf 4.1 Gauge fields and ghosts}

Contributions from the 
gauge fields and ghosts to the effective potential are given by 
\begin{equation}
V^{gauge + ghost}=-{i\over 2}{\rm tr~ln}D_L D^L(A_{y^I} ) ~~,
\label{potential1}
\end{equation}
where 
$D_L D^L( A_{y^I} )=\partial_{\mu}\partial^{\mu}
-\sum_{I=1}^2 D_{y^I}^2( A_{y^I} )$.
One needs to find  eigenvalues of the 
mass operator $D_{y^I}^2( A_{y^I} )$ to evaluate the effective
potential.

With (\ref{BC7}), the parity assignment for  for $A_{y^I}$ is given by
\begin{eqnarray} 
A_{y^I}^{a=1, 2}&:& (P_0, P_1, P_2)=(+++) ~,\\
A_{y^I}^{a=3}   &:& (P_0, P_1, P_2)=(---) ~.
\label{componentBC1}
\end{eqnarray}
The mass operator for $A_{y^I}^a$ ($a=1,2,3$) for  the background field
configuration (\ref{wilson6}) is obtained by inserting the mode expansion
(\ref{mode1}).  Due to the nonvanishing background $(\alpha,\beta)$, 
$A_{y^I}^2$ and $A_{y^I}^3$ mix with each other.
It is given by
\beqn
&&\hskip -1cm
\sum_{I=1}^{2}D_{y^I}^2=
{1\over R_1^2}
\pmatrix{
n^2 & 0 & 0                \cr
0 & n^2+\alpha^2 & 2n\alpha \cr
0 & 2n\alpha & n^2+\alpha^2  \cr}
+{1\over R_2^2}
\pmatrix{
m^2 & 0 & 0                \cr
0 & m^2+\beta^2 & 2m\beta \cr
0 & 2m\beta & m^2+\beta^2  \cr}  \cr
\noalign{\kern 10pt}
&&\hskip 8cm
\hbox{for } (n,m) \in K_+ ~~.
\label{mass-gauge1}
\eeqn
The eigenvalues of the operator for $(n, m)\neq (0, 0)$ are easily 
obtained as
\begin{equation}
\left({n\over R_1}\right)^2+\left({m\over R_2}\right)^2,\quad 
\left({{n\pm\alpha}\over R_1}\right)^2
   +\left({{m\pm\beta}\over R_2}\right)^2  \quad
\hbox{for } (n,m) \in K_+~.
\label{mass-gauge2}
\end{equation}
Zero modes $(n,m)=(0,0)$ exist only for $A_{y^I}^{a=1, 2}$.
Eigenvalues for the zero modes are given by
\begin{equation}
0, \quad \left({\alpha\over R_1}\right)^2
   +\left({\beta \over R_2}\right)^2  ~~.
\label{mass-gauge3}
\end{equation}

In a similar way, we can compute  
contributions from $A_{\mu}^{a=1, 2, 3}$ to the effective potential. In this
case the parity assignment is 
\begin{eqnarray} 
A_{\mu}^{a=1, 2}&:& (P_0, P_1, P_2)=(---),\\
A_{\mu}^{a=3}   &:& (P_0, P_1, P_2)=(+++).
\end{eqnarray}
The mass matrix has the same structure as before. 
Only $a=3$ component of $A_{\mu}^a$ has a zero mode. Hence eigenvalues of
the mass operator are
\beqn
&&\hskip -1cm
\left({n\over R_1}\right)^2+\left({m\over R_2}\right)^2,\quad 
\left({{n\pm\alpha}\over R_1}\right)^2
   +\left({{m\pm\beta}\over R_2}\right)^2  \quad
\hbox{for } (n,m) \in K_+~  \cr
\noalign{\kern 10pt}
&&\hskip -1cm
 \left({\alpha\over R_1}\right)^2
   +\left({\beta \over R_2}\right)^2  ~~.
\label{mass-gauge4}
\eeqn
The mass matrix for ghost fields is the same as that for $A_{\mu}$.
Contributions to the effective potential from $A_{\mu}$ and ghosts are,
therefore, $4-2=2$ times contributions coming from the spectrum 
(\ref{mass-gauge4}).  

In six dimensions there are two extra-dimensional components $A_{y^I}$.
Therefore, if one adds (\ref{mass-gauge2}),  (\ref{mass-gauge3}) and
(\ref{mass-gauge4}), one obtains two copies of
\beqn
&&\hskip -1cm
\left({n\over R_1}\right)^2+\left({m\over R_2}\right)^2 ~~,~~
\left({{n +\alpha}\over R_1}\right)^2
   +\left({{m +\beta}\over R_2}\right)^2  ~~,~~
\left({{n -\alpha}\over R_1}\right)^2
   +\left({{m -\beta}\over R_2}\right)^2  ~~, \cr
\noalign{\kern 10pt}
&&\hskip 4cm
(-\infty < n,m < +\infty ) ~~,
\label{mass-gauge5}
\eeqn
for the mass spectrum.  Here we used the fact that $K_+$ covers
a half of the integer lattice plane after $(0,0)$ is removed.

\begin{figure}[tbh]
\centering
\leavevmode
\includegraphics[width=9.cm]{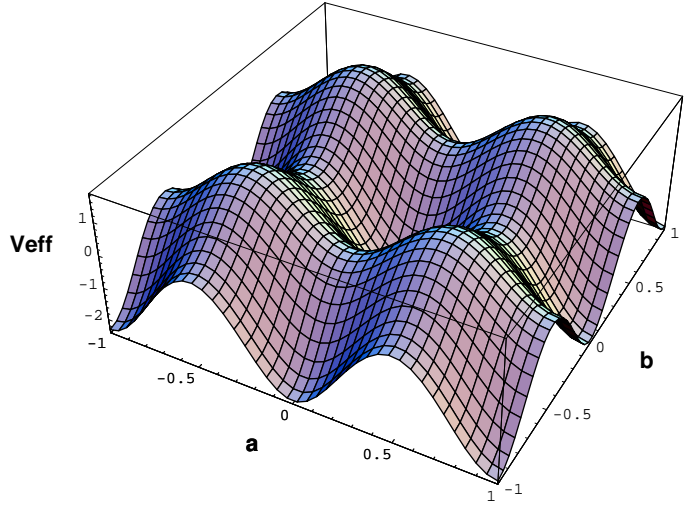}
\caption{The effective potential $V_\eff(\alpha, \beta)$,
(\ref{potentail-gauge3}),  in the  pure gauge theory with $R_1=R_2$. 
There are four degenerate minima at $(\alpha,\beta)=(0,0)$, 
(1,0), (0,1), and (1.1). All of them correspond to $U(1)$ symmetric
states.}
\label{fig-gauge-ghost}
\end{figure}

The contributions from the gauge fields and ghost fields are summarized as
\beqn
&&\hskip -1cm
V_\eff(\alpha,\beta)^{gauge} \cr
\noalign{\kern 10pt}
&&\hskip -1cm 
=2 \cdot {1\over 2} \int{{d^4p_E}\over{(2\pi)^4}} \,
{{1}\over{2\pi^2 R_1 R_2}}    
 \sum_{n=-\infty}^{\infty}\sum_{m=-\infty}^{\infty}
\Bigg\{ 
2 \ln 
\bigg[ p_E^2+ \bigg( {n+\alpha \over R_1} \bigg)^2
+\bigg( {{m+\beta}\over R_2} \bigg)^2  \bigg]  \cr
\noalign{\kern 10pt}
&&\hskip 6cm 
+ \ln \bigg[ p_E^2+\left({n\over R_1}\right)^2
    +\left({m\over R_2}\right)^2 \bigg] \Bigg\} ~~.
\label{potential-gauge1}
\eeqn  
Here the Wick rotation has been made and $p_E$ stands for the Euclidean 
momenta in four dimensions. As shown in 
refs.\ \cite{LeeHo, gaugeHiggs1}
\beqn
&&\hskip -1cm
I(\alpha, \beta)\equiv 
{1\over 2} \int{{d^4p_E}\over{(2\pi)^4}} \, 
{{1}\over{2\pi^2 R_1 R_2}}  
\sum_{n=-\infty}^{\infty}\sum_{m=-\infty}^{\infty}
{\rm ln}
\Bigl[p_E^2+\left({{n+\alpha}\over R_1}\right)^2
+\left({{m+\beta}\over R_2}\right)^2
\Bigr]  \cr
\noalign{\kern 10pt}
&&\hskip -0cm 
=- {{1}\over{16\pi^9}}
\Bigg\{ 
{1\over R_1^6}\sum_{n=1}^{\infty}{{\cos(2\pi n \alpha)}\over n^6}
+{1\over R_2^6}\sum_{m=1}^{\infty}
{{\cos(2\pi m \beta)}\over m^6}  \cr
\noalign{\kern 10pt}
&&\hskip 1.cm 
+2\sum_{n=1}^{\infty}\sum_{m=1}^{\infty}
{{\cos(2\pi n \alpha)\cos(2\pi m\beta)}
\over {(n^2R_1^2 + m^2 R_2^2)^3}}  \Bigg\}
+(\alpha, \beta~\mbox{-independent~~terms}) ~~.
\label{potential-gauge2}
\eeqn
In terms of $I(\alpha,\beta)$,  
\begin{equation}
V_\eff(\alpha,\beta)^{gauge}  
= 4 I(\alpha,\beta) + 2 I(0,0) ~~,
\label{potentail-gauge3}
\end{equation}
which is depicted in fig.\  \ref{fig-gauge-ghost}.
We note that one unit of $I$ represents contributions to the effective
potential from two physical degrees of freedom on $M^4 \times (T^2/Z_2)$.

\vskip .5cm
\leftline{\bf 4.2 Scalar fields in the fundamental representation}

A scalar field $H(x,y)=  (H_1 , H_2)^t$ in the fundamental
representation satisfies (\ref{BCs1}), or 
\beeq
H(x, \vec z_j - \vec y)=\eta_j^s \, \tau^3 \, H(x, \vec z_j + \vec y) 
\quad (j=0,1,2,3)~~.
\label{BCs3}
\eneq
Each component of $H$ is a $Z_2$ singlet. 
The parity assignment is
\beqn
H_1: &&(P_0,P_1,P_2) =(+\eta_0, +\eta_1, +\eta_2) \cr
H_2: &&(P_0,P_1,P_2) =(-\eta_0, -\eta_1, -\eta_2) ~~.
\label{parity-s1}
\eeqn
Consequently the mode expansion of the doublet $H$ is given by one of
pairs in (\ref{mode1}) - (\ref{mode4}).  

Let us first examine the case $\eta_0=\eta_1=\eta_2=+1$ or $-1$.  The
mode expansion of $(H_1,H_2)$ is given by a pair in (\ref{mode1}).
When the mass operator
\[
\sum_{I=1}^{2}D_{y^I}^2 = 
\pmatrix{\dd_{y^1} & - i\alpha/2R_1 \cr
          - i\alpha/2R_1 & \dd_{y^1}   }^2
+ \pmatrix{\dd_{y^2} & - i\beta/2R_2 \cr
          - i\beta/2R_2 & \dd_{y^2}   }^2
\]
acts on $(n,m)$  $(\in K_+)$ components in the mode expansion of $H$, it yields
a matrix
\beeq 
{1\over R_1^2}
\pmatrix{
n^2 +{1\over 4} \alpha^2  & i\alpha n \cr
-i\alpha n  & n^2+ {1\over 4}\alpha^2 \cr} 
+{1\over R_2^2}
\pmatrix{
m^2+ {1\over 4} \beta^2   & i\beta m \cr
-i\beta m  & m^2+ {1\over 4}\beta^2 \cr}  ~~,
\label{spectrum-s1}
\eneq
which has eigenvalues
\begin{equation}
{(n + \onehalf \alpha)^2 \over R_1^2}
+ {(m + \onehalf \beta)^2 \over R_2^2} ~~,~~
{(n - \onehalf \alpha)^2 \over R_1^2}
+ {(m - \onehalf \beta)^2 \over R_2^2} 
\hskip .5cm \hbox{where } (n,m)\in K_+ ~.
\label{spectrum-s2}
\end{equation}
Only one of $H_1$ or $H_2$ has a zero mode ($y^I$-independent mode).
Its eigenvalue for $\sum_{I=1}^{2}D_{y^I}^2$ is
\beeq
{\alpha^2 \over 4R_1^2}+ { \beta^2 \over 4R_2^2} ~~.
\label{spectrum-s3}
\eneq
Combining (\ref{spectrum-s2}) and (\ref{spectrum-s3}), one obtains
\beeq
{(n + \onehalf \alpha)^2 \over R_1^2}
+ {(m + \onehalf \beta)^2 \over R_2^2} \qquad 
(-\infty < n,m < +\infty)~.
\label{spectrum-s4}
\eneq

Analysis in other cases of parity assignment $(\eta_0, \eta_1,\eta_2)$ 
is almost the same.  The mode expansion is given by one of  pairs in 
(\ref{mode2}) - (\ref{mode4}).  There is no zero mode.
At this junction it is convenient to introduce $\delta_j$ by
\beqn
&&\hskip -1cm
\delta(\eta) = \cases{0 &for $\eta=+1$ ~, \cr
                       1 &for $\eta= -1$ ~,\cr}  \cr
\noalign{\kern 10pt}
&&\hskip -1cm
\delta_j = \delta(\eta_0 \eta_j) \quad (j=1,2) ~.
\label{delta-eta}
\eeqn
The only change arising when the mass operator
$\sum_{I=1}^{2}D_{y^I}^2$ acts on $(n,m)$ components in the mode expansion 
is that $n$ and $m$ in the matrix (\ref{spectrum-s1}) are replaced by
$n+\onehalf \delta_1$ and $m+\onehalf \delta_2$, respectively.
Consequently eigenvalues of $\sum_{I=1}^{2}D_{y^I}^2$ are given by
\beeq
{[ n + \onehalf (\alpha +\delta_1) ]^2 \over R_1^2}
+ {[m + \onehalf (\beta + \delta_2)]^2 \over R_2^2} \qquad 
(-\infty < n,m < +\infty) 
\label{spectrum-s5}
\eneq
in all cases.

Contributions of one scalar doublet to the effective potential is 
found, from (\ref{spectrum-s5}) to be
\beeq
V_\eff(\alpha, \beta)^{sF} 
= 2 I[\onehalf (\alpha +\delta_1), \onehalf (\beta + \delta_2)] ~~.
\label{potential-sF1}
\eneq
Here the factor 2 accounts for the complex nature of the field $H$.

\vskip .5cm
\leftline{\bf 4.3 Weyl fermions in the fundamental representation}

Let us next consider contributions to the effective potential from 
fermions in the fundamental representation.  We start with a Weyl
fermion  satisfying $\Gamma^7\psi=-\psi$ and take all the sign 
factor $(\eta_0, \eta_1, \eta_2)=(+,+,+)$. Then, the
mode expansion  with the boundary condition (\ref{BCf1}) or (\ref{BCf2}) 
 is given  by
\begin{eqnarray}
&&\hskip -1cm
\pmatrix{U_{L1} \cr U_{L2}} (x, y^I) =
{1\over \sqrt{2\pi^2 R_1R_2} }
\pmatrix{U_{L1 (0 0)} (x) \cr 0} \cr
\noalign{\kern 10pt}
&&\hskip 1cm
 +{1\over \sqrt{\pi^2 R_1 R_2} }
\sum_{(n,m) \in K_+}
 \pmatrix{ U_{L1(n m)}(x)  \cr U_{L2 (n m)}(x) }
 \pmatrix{\cos \cr \sin}
\left({n\over R_1}y^1+{m\over R_2}y^2\right) ~,\cr
\noalign{\kern 10pt}
&&\hskip -1cm
\pmatrix{D_{R1} \cr D_{R2}} (x, y^I)=
{1\over \sqrt{2\pi^2 R_1R_2} }
\pmatrix{0 \cr D_{R2 (0 0)} (x)} \cr
\noalign{\kern 10pt}
&&\hskip 1cm
+{1\over \sqrt{\pi^2 R_1 R_2} }
\sum_{(n,m) \in K_+}
 \pmatrix{ D_{R1(n m)}(x)  \cr D_{R2 (n m)}(x) }
 \pmatrix{\sin \cr \cos }
\left({n\over R_1}y^1+{m\over R_2}y^2\right) ~.
\label{mode-f1}
\end{eqnarray}
Note that each of $U_{La}$ or $D_{Ra}$ is a four-component
spinor with definite four-dimensional  chirality.
The mass operator for $U_{La}$ in this basis is given by
\begin{equation} 
\sum_{I=1}^{2}D_{y^I}^2  \Rightarrow
{1\over R_1^2}
\pmatrix{
n^2 +{1\over 4} \alpha^2& i\alpha n \cr
-i\alpha n & n^2+{1\over 4} \alpha^2 \cr}
+{1\over R_2^2}
\pmatrix{
m^2+{1\over 4} \beta^2& i\beta m \cr
-i\beta m & m^2+{1\over 4}\beta^2 \cr}
\label{mass-f1}
\end{equation}
for $(n,m) \in K_+$.
Eigenvalues  are given by
\begin{equation}
\left({{n + {1 \over 2} \alpha}\over R_1}\right)^2
+\left({{m + {1\over 2}\beta }\over R_2}\right)^2  ~~,~~
\left({{n - {1\over 2}\alpha}\over R_1}\right)^2
+\left({{m - {1\over 2}\beta}\over R_2}\right)^2 ~~
\hbox{for~}(n,m) \in K_+
\label{spectrum-f1}
\end{equation}
and 
\begin{equation}
\left({\alpha\over 2R_1}\right)^2
+\left({\beta\over 2R_2}\right)^2
\label{spectrum-f2}
\end{equation}
for the zero mode $U_{L1(00)}$.  Combining (\ref{spectrum-f1})
and (\ref{spectrum-f2}), one finds
\beeq
\left({{n + {1 \over 2} \alpha}\over R_1}\right)^2
+\left({{m + {1\over 2}\beta }\over R_2}\right)^2  
\qquad (-\infty < n,m < + \infty) ~.
\label{spectrum-f3}
\eneq
The spectrum is the same as for a scalar field in the fundamental
representation.

Eigenvalues of the mass operator for $D_{Ra}$ are the 
same as those for $U_{La}$. 
Therefore the contributions to the effective potential from 
a Weyl fermion in the fundamental representation 
with $(\eta_0,\eta_1,\eta_2)=(+,+,+)$ is
given by
\begin{equation}
V^{fF}= - 4 I({\alpha\over 2}, {\beta\over 2}) ~~.
\label{potential-f1}
\end{equation}
The minus sign is due to fermi statistics.

Extension to other cases of  $(\eta_0,\eta_1,\eta_2)$ is
straightforward.  The basis for the mode expansion (\ref{mode-f1}),
which corresponds to (\ref{mode1}), is changed to one of 
(\ref{mode2}) - (\ref{mode4}).  The resultant spectrum of the
mass operator $\sum_{I=1}^{2}D_{y^I}^2$ is the same as in the scalar
field case.  $(n,m)$ in (\ref{spectrum-f3}) is replaced by
$(n+ \onehalf \delta_1,m + \onehalf \delta_2)$ where $\delta_j$ is
defined in (\ref{delta-eta}).  Consequently the  contributions to the
effective potential from  a Weyl fermion in the fundamental
representation is summarized as
\beeq
V^{fF}= - 4 I({\alpha + \delta_1\over 2}, {\beta + \delta_2\over 2}) ~~.
\label{potential-f2}
\eneq

\vskip .5cm
\leftline{\bf 4.4 Weyl fermions and scalars in the adjoint
representation}

Contributions of matter fields in the adjoint representation are 
easily obtained as in the preceding subsections.   
Consider a Weyl fermion.   Note that 
$D_M \psi = \dd_M \psi + i g[A_M, \psi]$.  With the background
fields (\ref{wilson6}) 
\beqn
&&\hskip -1cm
2 \Tr \psibar i( \Gamma^5 D_5 + \Gamma^6  D_6)  \psi 
=\psibar {}^1 i ( \Gamma^5 \dd_{y^1} + \Gamma^6  \dd_{y^2})  \psi^1 \cr
\noalign{\kern 10pt}
&&\hskip -.5cm
+ (\psibar {}^2, \psibar {}^3) i \Bigg\{
\Gamma^5 \pmatrix{\dd_{y^1} & \alpha/R_1 \cr
                 -\alpha/R_1 & \dd_{y^1} }
+ \Gamma^6 \pmatrix{\dd_{y^2} & \beta/R_2 \cr
                 -\beta/R_2 & \dd_{y^2} }    \Bigg\} 
\pmatrix{\psi^2 \cr \psi^3} ~.
\label{lagrangian-fAd}
\eeqn
The parity assignment for a Weyl fermion
satisfying $\Gamma^7\psi=-\psi$ is
\begin{eqnarray}
\pmatrix{U_{L}^{a=1} \cr D_{R}^{a=1}} ,
\pmatrix{U_{L}^{a= 2} \cr D_{R}^{a= 2}}
&:& (P_0, P_1, P_2)
= \cases{(-\eta_0, -\eta_1, -\eta_2) \cr
         (+\eta_0, +\eta_1, +\eta_2) \cr}  ~, \cr
\noalign{\kern 10pt}
\pmatrix{U_{L}^{a= 3} \cr D_{R}^{a= 3}} \hskip 1.0cm
&:& (P_0, P_1, P_2)
= \cases{(+\eta_0, +\eta_1, +\eta_2)  \cr
         (-\eta_0, -\eta_1, -\eta_2)\cr}    ~.
\label{parity-fAd}
\end{eqnarray}

The net consequence in the evaluation of the mass operator 
$\sum_{I=1}^{2}D_{y^I}^2$ is that $(\alpha,\beta)$ in the case
of fermions in the fundamental representation is replaced by
$(2\alpha,2\beta)$.  Contributions to the effective potential
is summarized as
\beeq
V^{f,Ad}
= - 2 \Big\{ I (\onehalf \delta_1, \onehalf \delta_2) +
2 I(\alpha + \onehalf \delta_1, \beta + \onehalf\delta_2) \Big\}
~~.
\label{potential-fAd1}
\eneq
Similarly, for a real scalar field in the adjoint representation
we have
\beeq
V^{s,Ad}
=  {1\over 2} \Big\{ I (\onehalf \delta_1, \onehalf \delta_2) +
2 I(\alpha + \onehalf \delta_1, \beta + \onehalf\delta_2) \Big\}
~~.
\label{potential-sAd1}
\eneq

Adding  contributions from  gauge fields and ghosts, we 
immediately see that $V^{gauge + ghost}+
V^{f,Ad}=0$ if $\delta_1=\delta_2=0$. This is because
in six dimensions $(A_M, \psi_{adj})$ forms the vector multiplet 
of ${\cal N}=1$ supersymmetry \cite{bss} and their on-shell degrees of freedom
are equal to each other. 
Therefore, the contributions from bosons and fermions are canceled 
to yield the vanishing effective potential. It is important to 
observe that the cancellation holds  only for the sign 
assignment $(\eta_0, \eta_1, \eta_2)=(+++)$ or $(---)$. For the other
cases, the  effective potential does not vanish.
The ${\cal N}=1$ supersymmetry is broken by the
different assignment of the sign factors $\eta_i$ for bosons and 
fermions. This is similar to the Scherk-Schwarz breaking of 
supersymmetry \cite{SS}, in which  different boundary 
conditions for bosons and 
fermions are imposed.

\vskip .5cm
\leftline{\bf 4.5 $Z_2$ doublets}

Twists along  noncontractible loops on $T^2$ can be introduced for each field
by doubling the number of degrees of freedom.  As we see below, these
$T^2$ twists give   fermions additional masses in four dimensions.  This
may be very important in the  phenomenological viewpoint, as these twists
can substitute Yukawa interactions.  We prepare a pair of Weyl fermions
$(\psi, \psi')$ satisfying
\beqn
&&\hskip -1cm 
\pmatrix{\psi \cr \psi'} (x,  - \vec y) 
= \eta_0  \, T[P_0] \,
(i\Gamma^4\Gamma^5)
\pmatrix{\psi \cr - \psi'}  (x,  \vec y) ~~, \cr
\noalign{\kern 10pt}
&&\hskip -1cm 
\pmatrix{\psi \cr \psi'} (x, y^1 + 2\pi R_1, y^2) 
 = \pmatrix{ \cos \pi a & - \sin \pi a \cr \sin \pi a & \cos \pi a}
\eta_0 \eta_1 \, T[U_1] 
  \pmatrix{\psi \cr \psi'} (x, \vec y)  ~~, \cr
\noalign{\kern 10pt}
&&\hskip -1cm 
\pmatrix{\psi \cr \psi'} (x, y^1, y^2 + 2\pi R_2) 
 = \pmatrix{ \cos \pi b & - \sin \pi b \cr \sin \pi b & \cos \pi b}
\eta_0 \eta_2 \, T[U_2] 
  \pmatrix{\psi \cr \psi'} (x, \vec y)  ~~.
\label{BCdoublet1}
\eeqn
Nonvanishing $a$ and $b$ give twists on the pair $(\psi, \psi')$.  
Note that each pair can have its own $(a,b)$.

Let us illustrate it by  considering fermions in the fundamental
representation for which $ T[P_0] \, \psi = P_0  \psi$ etc..
Take $\eta_0=1$, $P_0= P_1= P_2= \tau^3$, $U_1=U_2= \Itwo$.   With the notation
in (\ref{updown1}), $(U_a, U_a')$ and $(D_a, D_a')$ ($a=1,2$) form
$Z_2$ doublets.   Their mode expansions are given, as in
(\ref{modeD1}),  by 
\beqn
&&\hskip -1cm 
\pmatrix{U_{R1} \cr U_{R1}'} (x, \vec y) 
=  {1\over \sqrt{2 \pi^2 R_1 R_2}} 
\sum_{n,m=-\infty}^\infty   \hat U_{R1,nm} (x) 
\pmatrix{\cos z_{nm}(\vec y \,)\cr \sin z_{nm}(\vec y \,)}  \cr
\noalign{\kern 10pt}
&&\hskip -1cm 
\pmatrix{U_{R2} \cr U_{R2}'} (x, \vec y) 
=  {1\over \sqrt{2 \pi^2 R_1 R_2}} 
\sum_{n,m=-\infty}^\infty   \hat U_{R2,nm} (x) 
\pmatrix{- \sin z_{nm}(\vec y \,)\cr \cos z_{nm}(\vec y \,)}  \cr
\noalign{\kern 10pt}
&&\hskip 1cm
z_{nm}(\vec y \,)
 = \bigg( n+ \myfrac{a + \delta_1}{2} \bigg) \myfrac{y^1}{R_1}
 + \bigg( m+ \myfrac{b + \delta_2}{2} \bigg) \myfrac{y^2}{R_2} ~~.
\label{modeD2}
\eeqn
Similar expansions hold for $D_{La}$ and $D_{La}'$ as well.

The nonvanishing Wilson line phases $\alpha$ and $\beta$ mix
$ \hat U_{R1,nm}$ and  $\hat U_{R2,nm}$ as in the subsection 4.3.  
The resultant mass matrix takes the same form as in (\ref{mass-f1})
where $n$ and $m$ are replaced by 
$n+\onehalf(a+\delta_1)$ and $ m+\onehalf (b+\delta_2)$, respectively. 
Hence eigenvalues are given by
\beqn
&&\hskip -1cm
{1\over R_1^2} \Bigg( n+ \myfrac{a + \delta_1 + \alpha}{2} \Bigg)^2
+ {1\over R_2^2} \Bigg( m+ \myfrac{b + \delta_2 + \beta}{2} \Bigg)^2 ~~, \cr
\noalign{\kern 10pt}
&&\hskip -1cm
{1\over R_1^2} \Bigg( n+ \myfrac{a + \delta_1 - \alpha}{2} \Bigg)^2
+ {1\over R_2^2} \Bigg( m+ \myfrac{b + \delta_2 - \beta}{2} \Bigg)^2 ~~, \cr
\noalign{\kern 10pt}
&&\hskip 1cm
-\infty < n , m < + \infty ~~.
\label{spectrum-doublet1}
\eeqn
To summarize, the contributions to the effective potential from each $Z_2$
doublet of Weyl fermions in the fundamental representation are given by
\beeq
V^{fF}_{\rm doublet} = 
-4 \bigg\{ I \Big[ \onehalf (\alpha + a + \delta_1), 
                   \onehalf ( \beta + b + \delta_2) \Big]
 + I \Big[ \onehalf (\alpha - a + \delta_1), 
                   \onehalf ( \beta - b + \delta_2) \Big] \bigg\} ~~.
\label{potential-doublet1}
\eneq
Extension to fields in other representation is straightforward.

\vskip .5cm
\leftline{\bf 4.6 The total effective potential}

Adding all the contributions of $Z_2$ singlet fields, we find that the
total effective potential for the Wilson line phases is given by
\beqn
&&\hskip -1cm
V_\eff (\alpha, \beta) =
4 I(\alpha, \beta) + 2 I(0,0) \cr
\noalign{\kern 10pt}
&&\hskip -.5cm
+ \sum_{\delta_1,\delta_2}
 2 \Big\{ N^{s,F}_{(\delta_1\delta_2)} 
    - 2 N^{f,F}_{(\delta_1\delta_2)}  \Big\} \, 
  I[\onehalf(\alpha + \delta_1), \onehalf (\beta +\delta_2)] \cr
\noalign{\kern 10pt}
&&\hskip -.5cm
+ \sum_{\delta_1,\delta_2}
 {1\over 2} \Big\{ N^{s,Ad}_{(\delta_1\delta_2)} 
    - 4 N^{f,Ad}_{(\delta_1\delta_2)}  \Big\} \, 
\Big\{ I (\onehalf \delta_1, \onehalf \delta_2) +
2 I(\alpha + \onehalf \delta_1, \beta + \onehalf\delta_2) \Big\} ~.
\label{potential-all1}
\eeqn
Here $N^{f,F}_{(\delta_1\delta_2)}$ and $N^{f,Ad}_{(\delta_1\delta_2)}$
are the numbers of Weyl fermion multiplets in the fundamental
and adjoint representation with the parity assignment
$(\delta_1\delta_2)$, respectively.
 $N^{s,F}_{(\delta_1\delta_2)}$ and $N^{s,Ad}_{(\delta_1\delta_2)}$
are defined similarly for scalar fields.  
($N^{s,Ad}_{(\delta_1\delta_2)}$ counts the number of real scalar field
multiplets.)  If there exist fields of $Z_2$ doublets, their contributions
need to be added.

The true vacuum is given by 
the global minimum of (\ref{potential-all1}). As we see in the 
following section, the global minimum can be located at nonvanishing
$(\alpha, \beta)$.

\ignore{
\begin{figure}[tbh]
\centering
\leavevmode
\includegraphics[width=9.cm]{scalar.eps}
\caption{Noda's scalar.eps .  Contribution to $V_\eff(\alpha,\beta)$
from a scalar field in the fundamental representation with 
$(\delta_1,\delta_2)=(0,0)$.}
\label{scalar}
\end{figure}
}

\sxn{Gauge symmetry breaking}

The true vacuum is determined by the global minimum of 
 the effective potential for the Wilson line phases 
(\ref{potential-all1}).  We recall that $\alpha$ and $\beta$ are
phase variables with a period 2.  The function $I(\alpha, \beta)$,
which is defined in (\ref{potential-gauge2}), satisfies
$I(\alpha+1, \beta) = I(\alpha, \beta+1)=I(\alpha,\beta)$.  It has
the global minimum at $(0,0)$, the global maximum at $(\onehalf,
\onehalf)$, and saddle points at $(\onehalf, 0)$ and $(0, \onehalf)$
$(mod ~ 1)$,  respectively.

\vskip .5cm
\leftline{\bf 5.1 Pure gauge field theory}

The case of the pure $SU(2)$ gauge theory has been already examined in
section 4.1.  The effective  potential is given by
(\ref{potentail-gauge3}).
The configurations that minimize the
potential are found to be 
\begin{equation}
(\alpha, \beta)=(0, 0) ~,~ (1,1) ~,~ (1,0) ~,~ (1,1) ~.
\label{minimum1}
\end{equation}  
\ignore{
where we have assumed the equal size of the two extra 
dimensions $R_1=R_2$.  {\bf (Necessary?)} 
}
We have seen that the 
phases $\alpha, \beta$ are determined dynamically.     

Let us discuss the gauge symmetry at low energies. First of all, the Wilson
line for the parameterization is given by
\begin{equation}
W_1 ={\rm e}^{i\pi\alpha \tau^1},~~W_2 ={\rm e}^{i\pi\beta\tau^1}.
\end{equation} 
Let us move to a new gauge, in which $\langle A_{y^I}^{\prime}\rangle =0$, 
by a gauge transformation  
\beeq
\Omega (\vec y ~; \alpha, \beta) = \exp \Bigg\{ i \bigg( 
  \myfrac{\alpha y^1}{2 R_1} + \myfrac{\beta y^2}{2 R_2}
   \bigg) \tau^1 \Bigg\} ~~.
\label{largeGT3}
\eneq
Then, new parity
matrices in (\ref{newBC1})  become
\begin{equation}
P_0^{\prime}=\tau^3,\quad 
P_1^{\prime}={\rm e}^{i\pi\alpha\tau^1}\tau^3,\quad
P_2^{\prime}={\rm e}^{i\pi\beta\tau^1}\tau^3 ~.
\label{newparity1}
\end{equation}  
As we have discussed, generators commuting with the 
new $P_i^{\prime}$ $(i=0, 1, 2)$ form the symmetry algebra at low energies.
For $(\alpha, \beta)=(0, 0)$, we 
have $P_0^{\prime}=P_1^{\prime}=P_2^{\prime}=\tau^3$.  $\onehalf \tau^3$ 
commutes with all the $P_i^{\prime}$, so that the $U(1)$ symmetry 
survives  at low energies. The symmetry of  boundary conditions 
at the tree level is not broken even at the quantum level.  

Taking into account the periodicity of the effective 
potential, the configurations $(\alpha, \beta)=(1, 0), (0, 1), (1, 1)$ also 
give the vacuum configurations. These configurations are
physically equivalent with $(\alpha, \beta)=(0, 0)$. In order to
see that, let us consider $(\alpha, \beta)=(1, 0)$, for which we have
$P_0^{\prime}=\tau^3, P_1^{\prime}=-\tau^3, P_2^{\prime}=-\tau^3$. 
Again, $\tau^3/2$ commutes with these parity matrices, so that there is
$U(1)$ gauge symmetry at low energies. One can also confirm that the mass
spectrum on each vacuum is the same.  Indeed, masses 
for $A_{\mu}^{a=3}$ are given 
by $(n+\alpha)^2 R_1^{-2}+(m+\beta)^2 R_2^{-2}$. 
$A_{\mu (n,m)=(0,0)}^{a=3}$ becomes a massless mode corresponding to
the $U(1)$ gauge symmetry for the 
configuration $(\alpha, \beta)=(0, 0)$, while   
$A_{\mu (n,m)=(-1,0)}^{a=3}$ is a massless mode 
for $(\alpha, \beta)=(1, 0)$. Likewise, a massless mode for the  
$U(1)$ gauge symmetry is given by $A_{\mu (n,m)=(0,-1)}^{a=3}$ and
$A_{\mu (n,m)=(-1,-1)}^{a=3}$ 
for $(\alpha, \beta)=(0, 1)$ and $(\alpha, \beta)=(1, 1)$, respectively.
Hence, the vacuum configuration related by the periodicity of the potential
is physically equivalent to each other and the mass spectrum on
each vacuum is obtained by shifting the K-K modes by the same amount of 
the periodicity.

\vskip .5cm
\leftline{\bf 5.2 With fermions in the fundamental representation}

\begin{figure}[tbh]
\centering
\leavevmode
\includegraphics[width=9.cm]{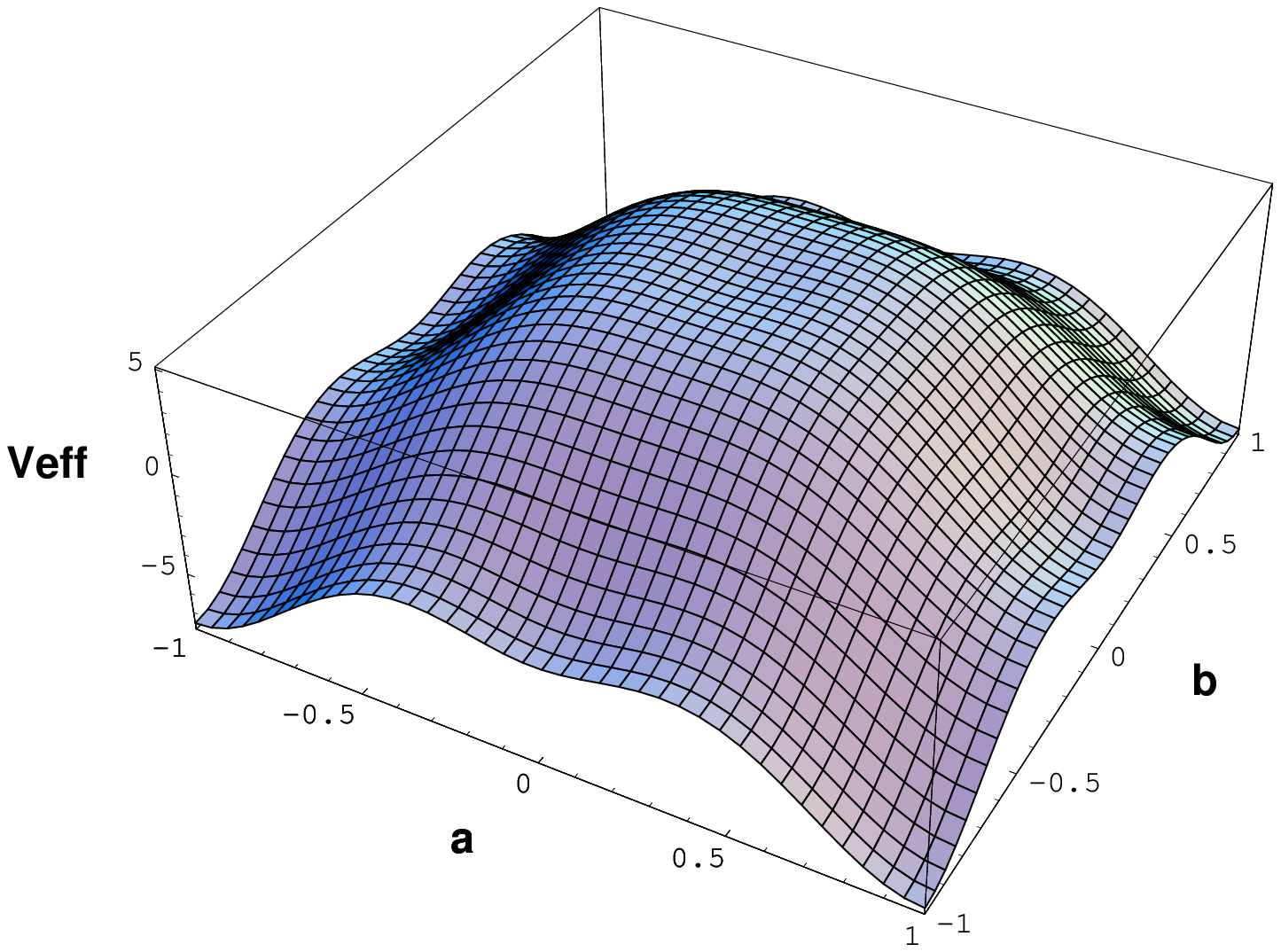}
\caption{$V_\eff(\alpha,\beta)$ for $N^{f,F}_{(00)}=3$ and $R_1=R_2$.}
\label{fig-fF}
\end{figure}

When there are additional fermions in the fundamental representation,
one of the configurations in  (\ref{minimum1}) becomes the 
global minimum of the effective potential.  Take, as an 
example, the case $N_{00}^{f,F} \not= 0$.  The potential becomes
\beeq
V_\eff (\alpha, \beta) =
4 I(\alpha, \beta) + 2 I(0,0) 
- 4 N^{f,F}_{00}  \,  I[\onehalf \alpha , \onehalf \beta ]  ~.
\label{potential-fF1}
\eneq
As $-I[\onehalf \alpha , \onehalf \beta ]$ takes the minimum value at
$(\alpha,\beta)=(1,1)$ $(mod ~ 2)$, the global minimum is located at 
$(\alpha,\beta)=(1,1)$.  
The physical symmetry is $U(1)$.  The effective potential for
$N^{f,F}_{(00)}=3$ is depicted in fig.\ \ref{fig-fF}.

If $N_{00}^{f,F} = 0$ and $N_{11}^{f,F} \not= 0$, the effective potential 
becomes
\beeq
V_\eff (\alpha, \beta) =
4 I(\alpha, \beta) + 2 I(0,0) 
- 4 N^{f,F}_{11}  \,  
  I[\onehalf \alpha + \onehalf , \onehalf \beta + \onehalf ]  ~.
\label{potential-fF2}
\eneq
In this case the global minimum is located at 
$(\alpha,\beta)=(0,0)$ $(mod ~ 2)$.

\vskip .5cm
\leftline{\bf 5.3 With fermions in the adjoint representation}

Let us consider the cases with fermions in the adjoint representation.
The effective potential is given by
\beqn
&&\hskip -1cm
V_\eff (\alpha, \beta) =
4 I(\alpha, \beta) + 2 I(0,0) \cr
\noalign{\kern 10pt}
&&\hskip -.5cm
- \sum_{\delta_1,\delta_2}
     2 N^{f,Ad}_{(\delta_1\delta_2)}  \, 
\Big\{ I (\onehalf \delta_1, \onehalf \delta_2) +
2 I(\alpha + \onehalf \delta_1, \beta + \onehalf\delta_2) \Big\} ~.
\label{potential-fAd2}
\eeqn
If only fermions with $(\delta_1\delta_2)=(00)$ exist, then
\beeq
V_\eff (\alpha, \beta) =
2(1 - N^{f,Ad}_{00})
\Big\{ 2I(\alpha, \beta) + I(0,0) \Big\} ~.
\label{potential-fAd3}
\eneq
For $N^{f,Ad}_{00} \ge 2$, the global minimum of the effective
potential is given by the global maximum of $I(\alpha, \beta)$. 
There are four degenerate minima located at $(\alpha, \beta)=
(\onehalf,\onehalf)$ $(mod ~1)$.

For the vacuum configuration $(\alpha, \beta)=(\onehalf,\onehalf)$, for
instance,  the new parity matrices in (\ref{newparity1}) 
are   given by 
\begin{equation}
P_0^{\prime}=\tau^3 ~,\quad 
P_1^{\prime}=\tau^2 ~,\quad
P_2^{\prime}=\tau^2 ~.
\label{newparity2}
\end{equation}  
There is no $SU(2)$ generator that commutes with all 
the $P_i^{\prime}$, so that the $U(1)$ gauge symmetry is broken.
As a result, there is no massless gauge boson. In fact, remembering that 
the mass spectrum for $A_{\mu (n,m)}^{a=3}$ is given 
by $(n+\alpha)^2 R_1^{-2}+(m+\beta)^2 R_2^{-2}$, for the
vacuum  configuration $(\alpha, \beta)=({1\over 2}, {1\over 2})$, we 
immediately see that none of $A_{\mu (n,m)}^{a=3}$ can be 
massless.  
There is no massless mode in $A_{\mu (n,m)}^{a=3}$ for 
non-integer values of $\alpha, \beta$ in general.

\vskip .5cm
\leftline{\bf 5.4 With $N^{f,F},  N^{f,Ad} \not= 0$}

In the examples described above, the configuration corresponding to
the global minimum of the effective potential is located at 
special points $(\alpha,\beta)=(0,0)$ where $\alpha$ and $\beta$ are
integers or half-odd-integers.
More generic configurations can be chosen if fermions in the 
fundamental representation and fermions in the adjoint representation 
coexist.

As an example let us examine the case with
 $N^{f,F}_{00} , N^{f,Ad}_{01} \not= 0$.
The effective potential is given by 
\beeq
V_\eff (\alpha, \beta) =
4 I(\alpha, \beta) 
    - 4 N^{f,F}_{00}   \, 
  I[\onehalf \alpha , \onehalf \beta ] 
    - 4 N^{f,Ad}_{01}  \, 
 I(\alpha , \beta + \onehalf )  ~~.
\label{potential-fFAd1}
\eneq
In the case $N^{f,F}_{00}=0$ the global minimum is located at
$(\alpha,\beta)=(0,0) ~(mod ~ 1)$ for $N^{f,Ad}_{01} \le 1$, while
at $(\alpha,\beta)=(\onehalf,0) ~(mod ~ 1)$ for $N^{f,Ad}_{01} \ge 2$.

\begin{figure}[tbh]
\centering
\leavevmode
\includegraphics[width=9.cm]{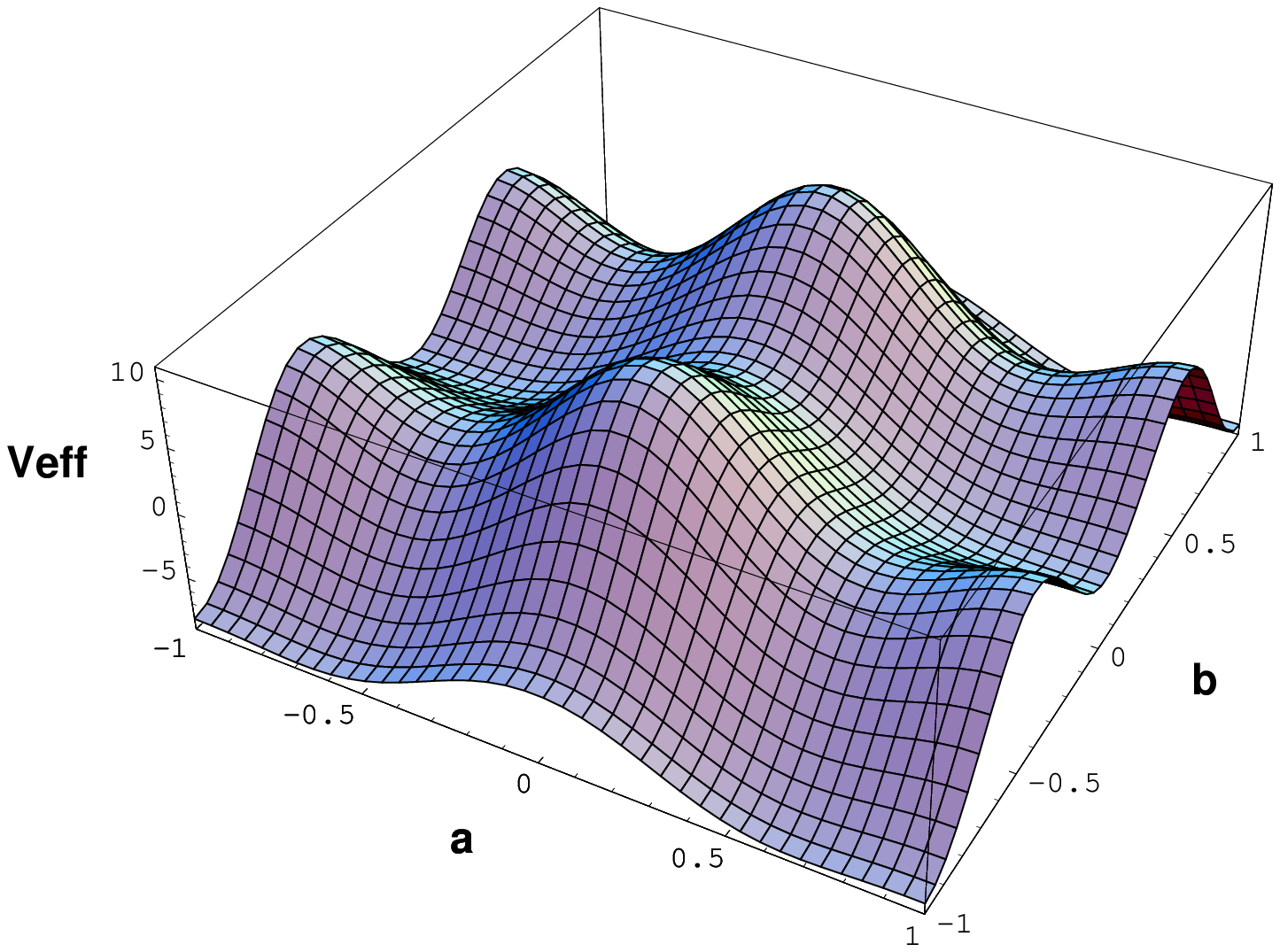}
\caption{$V_\eff(\alpha,\beta)$ for
$N^{f,F}_{(00)}= N^{f,Ad}_{(01)}=3$  with $R_1=R_2$. The global
minima are located at $(\alpha,\beta)=(\pm 0.678, 1) ~(mod ~ 2)$.}
\label{fig-fFAd1}
\end{figure}

\begin{figure}[tbh]
\centering
\leavevmode
\includegraphics[width=9.cm]{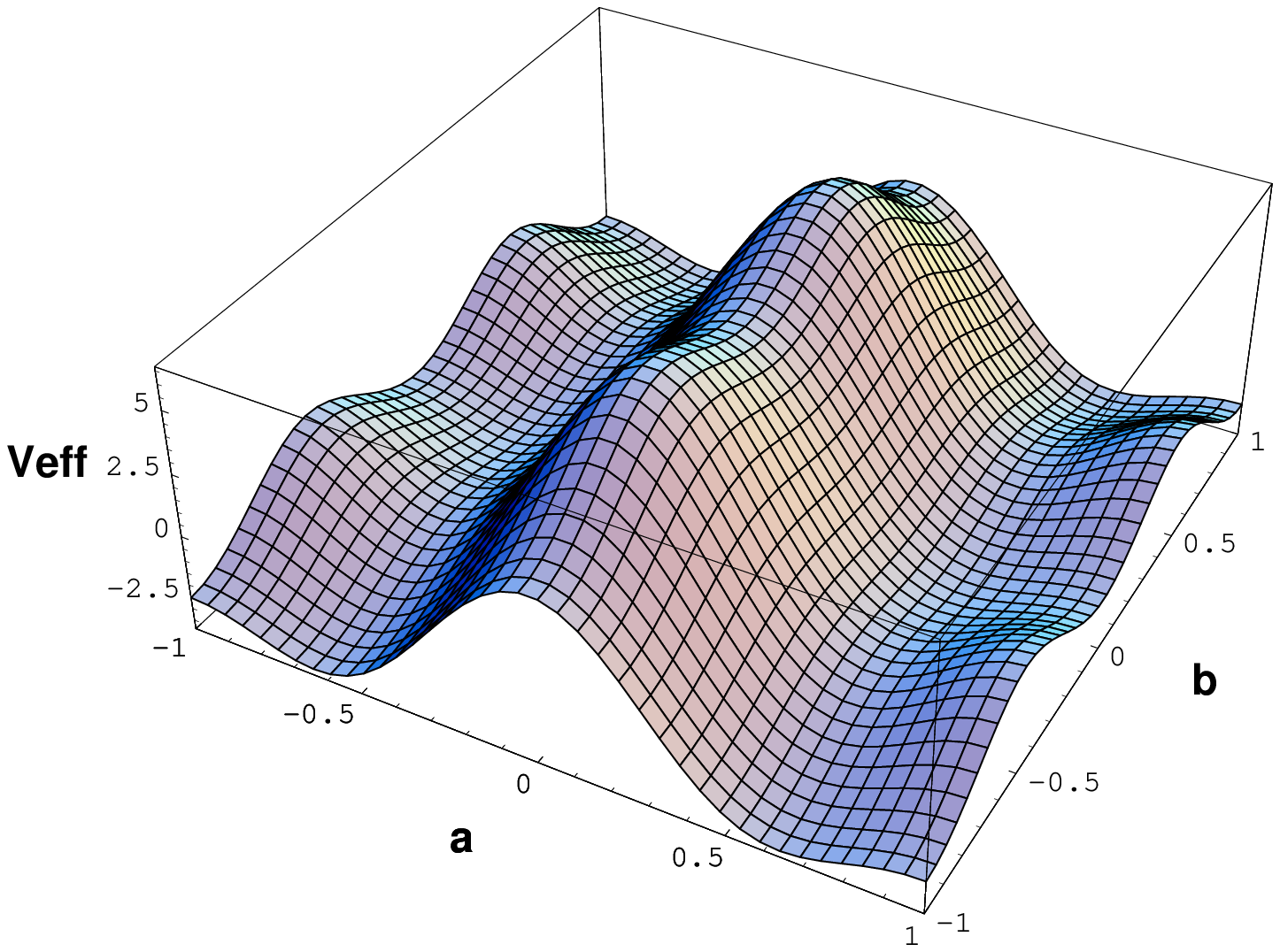}
\caption{$V_\eff(\alpha,\beta)$ for
$N^{f,F}_{(00)}= N^{f,Ad}_{(01)}=3$    with $1.3 R_1= R_2$. The global
minima are located at $(\alpha,\beta)=(\pm 0.636, 1) ~(mod ~ 2)$.}
\label{fig-fFAd2}
\end{figure}

Now add fermions in the fundamental representation with $N^{f,F}_{00} \not=
0$.  In the vicinity of $(\alpha,\beta)=(\onehalf,0) ~(mod ~ 1)$, 
$I[\onehalf \alpha , \onehalf \beta ]$ has a non-vanishing slope in the 
$\alpha$-direction.  Hence the location of the global minimum is shifted
in the $\alpha$-direction.  Furthermore, the four-fold degeneracy
existing in the case of  $N^{f,F}_{00} =0$ is partially lifted.  For instance,
the two degenerate global minima are located at 
 $(\alpha,\beta)=(\pm 0.555, 1) ~(mod ~ 2)$ for 
$(N^{f,F}_{00},N^{f,Ad}_{01}) =(1,3)$ with $R_1=R_2$.
For $(N^{f,F}_{00},N^{f,Ad}_{01}) =(3,3)$, 
the global minima are located at 
 $(\alpha,\beta)=(\pm 0.678, 1) ~(mod ~ 2)$ for $R_1=R_2$ and
$(\alpha,\beta)=(\pm 0.636, 1) ~(mod ~ 2)$ for $ R_2/R_1= 1.3$.
 See fig.\ \ref{fig-fFAd1}  and fig.\ \ref{fig-fFAd2}.
The minima are shifted to $(\alpha,\beta)=(\pm 0.600, 1) ~(mod ~ 2)$
for $(N^{f,F}_{00},N^{f,Ad}_{01}) =(3,4)$.

\vskip .5cm
\leftline{\bf 5.5 With fermions in $Z_2$-doublets}

It is of great interest from the phenomenological viewpoint to 
incorporate fermions in $Z_2$ doublets.  Intriguing models are obtained
if there are fermions in the adjoint representation 
($N^{f, Ad}_{00} , N^{f,Ad}_{01} \not= 0$) and fermions in $Z_2$ doublets in
the fundamental representation ($N^{f,F}_{00} \not= 0$) with the twist
parameters
$(a,b) \sim (0.5, -0.5)$.
The effective potential becomes 
\beqn
&&\hskip -1cm 
V_\eff (\alpha, \beta) =
4 I(\alpha, \beta)  
- 4 N^{f,Ad}_{00}  \,  I(\alpha , \beta ) 
- 4 N^{f,Ad}_{01}  \,  I(\alpha , \beta + \onehalf) \cr
\noalign{\kern 5pt}
&&\hskip .5cm
    - 4 N^{f,F}_{00, \rm doublet}   \, 
 \Big\{ \, I[\onehalf (\alpha + a) , \onehalf (\beta  + b)] 
   +  I[\onehalf (\alpha - a) , \onehalf (\beta  - b)] \, \Big\}  ~~.
\label{potential-fFAd2}
\eeqn

First take $(N^{f,Ad}_{00}, N^{f,Ad}_{01}) =(2, 0)$.
When $N^{f,F}_{00, \rm doublet} =0$, there are four degenerate global
minima at
$(\alpha, \beta) = (\pm\onehalf, \pm\onehalf)$ and
 $(\pm\onehalf, \mp\onehalf)$.  We add three generations of fermions
in the fundamental representation, $N^{f,F}_{00, \rm doublet} =3$.  
For $(a,b) = (\onehalf, -\onehalf)$, the degeneracy is partly lifted.
The effective potential has the global minima at 
$(\alpha, \beta) = (\pm\onehalf, \pm\onehalf)$.
Now we vary the values of $a$ and $b$.  For $(a,b) = (0.51, -0.51)$,
the global mimima move to $(\alpha, \beta) = (\pm 0.486, \pm 0.486)$.
For $(a,b) = (0.52, -0.52)$,
the global mimima move to $(\alpha, \beta) = (\pm 0.472, \pm 0.472)$.

As a second example, take $(N^{f,Ad}_{00}, N^{f,Ad}_{01}) =(0, 3)$.
When $N^{f,F}_{00, \rm doublet} =0$, there are four degenerate global
minima at $(\alpha, \beta) = (\pm\onehalf, 0)$ and $(\pm\onehalf, 1)$.  
Again we add three generations of fermions
in the fundamental representation, $N^{f,F}_{00, \rm doublet} =3$.  
For $(a,b) = (0.5, 0)$, the degeneracy is partly lifted.
The effective potential has the global minima at 
$(\alpha, \beta) = (\pm 0.5,1)$.  For $(a,b) = (0.5, 1)$, the global
minima are located at $(\alpha, \beta) = (\pm 0.5,0)$.
For $(a,b) = (0.52, 0)$, the global
minima are located at $(\alpha, \beta) = (\pm 0.479,1)$.

In all these cases the $SU(2)$ symmetry is completely broken.

\sxn{Mass generation}

As the Wilson line phases develop nonvanishing expectation 
values ($\alpha, \beta \not= 0$), the mass spectrum changes from
that at the tree level.  We are particularly interested in
the mass spectrum in four dimensions.

\vskip .5cm
\leftline{\bf 6.1 Four-dimensional gauge fields and  scalars}

Extra-dimensional components of gauge potentials $A^a_{y^I}$ play the 
role of four-dimensional Higgs scalar fields.   With the boundary 
condition (\ref{BC7}), the components $a=1,2$ of $A_{y^I}^{a}$ have
zero modes which serve as lower-dimensional scalars.  They are massless
at the tree level, but acquire nonvanishing masses at the quantum 
level. 

The fields $A^a_{y^I}$ acquire masses in two steps. When the global
minimum of the effective potential $V_\eff(\alpha, \beta)$ is located
at $(\alpha_\min , \beta_\min) \not= (0,0)$ $(mod ~ 2)$, the fields are
expanded around this configuration.  Through the gauge coupling
all fields in the four dimensions acquire masses of O($\alpha_\min/R_1$)
and of  O($\beta_\min/R_2$).  Some of $A^a_{y^I}$ may not be affected
by this correction, but they acquire nonvanishing masses from one loop
corrections.  It is a part of the Hosotani mechanism.\cite{YH2, HHHK}
The mechanism is similar to that of pseudo-Nambu-Goldstone
bosons and that of the little Higgs.\cite{lHiggs} 

The best way to understand this is to go to a new gauge in which
expectation values of Wilson line phases  vanish.
Perform a large gauge transformation 
$\Omega (\vec y~; \alpha_\min, \beta_\min)$ defined in (\ref{largeGT3}). 
In the new gauge $\la A_{y^I} \ra = 0$.  
The boundary conditions 
change to $(P_0, P_1, P_2) = (\tau^3, e^{i\pi \alpha_\min \tau^1} \tau^3,
e^{i\pi \beta_\min \tau^1} \tau^3)$ and 
$(U_1, U_2) = ( e^{i\pi \alpha_\min \tau^1} , e^{i\pi \beta_\min \tau^1})$. 

Let us look at the mass spectrum of four-dimensional gauge fields.
In this gauge $A_\mu^1 (x,\vec y)$ has a mode expansion of a
$Z_2$ singlet field with $(P_0, P_1, P_2) = (-,-,-)$ in (\ref{mode1}),
while $(A_\mu^3 (x,\vec y), A_\mu^2 (x,\vec y))$ forms a $Z_2$
doublet with $(a,b) = ( 2\alpha_\min,  2\beta_\min)$ in (\ref{modeD1}).
The spectrum is, therefore, 
\beqn
A_\mu^1 ~~ &:&
 \bigg( \myfrac{n}{R_1} \bigg)^2 + \bigg( \myfrac{m}{R_2} \bigg)^2
\hskip 3.cm  \hbox{where ~} (n,m) \in K_+ ~,  \cr
\noalign{\kern 10pt}
\pmatrix{A_\mu^3 \cr A_\mu^2} &:&
\bigg( \myfrac{n + \alpha_\min}{R_1} \bigg)^2 
  + \bigg( \myfrac{m + \beta_\min}{R_2} \bigg)^2
\qquad  \hbox{where ~} -\infty < n,m < + \infty ~.
\label{spectrumGauge1}
\eeqn
When $(\alpha_\min, \beta_\min) = (0,0)$, $A_\mu^1$ and $A_\mu^2$ have the
same spectrum and only $A_\mu^3$ has zero modes.
When $(\alpha_\min, \beta_\min) \not= (0,0)$, $A_\mu^2$ and $A_\mu^3$
mix to form mass eigenstates.  With this mixing in mind, it can be  said that
all three $SU(2)$ components of the gauge fields have distinct masses.

Similarly the spectrum of the extra-dimensional components $A_{y^I}^a$
is found.   $A_{y^I}^1 (x,\vec y)$ has a mode expansion of a
$Z_2$ singlet field with $(P_0, P_1, P_2) = (+,+,+)$ in (\ref{mode1}),
while $(A_{y^I}^2 (x,\vec y), A_{y^I}^3 (x,\vec y))$ forms a $Z_2$
doublet with $(a,b) = (- 2\alpha_\min, - 2\beta_\min)$ in (\ref{modeD1}).
The mass spectrum at the tree level is
\beqn
A_{y^I}^1 ~~ &:&
0 ~,~ \bigg( \myfrac{n}{R_1} \bigg)^2 + \bigg( \myfrac{m}{R_2} \bigg)^2
\hskip 2.3cm  \hbox{where ~} (n,m) \in K_+ ~,  \cr
\noalign{\kern 10pt}
\pmatrix{A_{y^I}^2 \cr A_{y^I}^3} &:&
\bigg( \myfrac{n - \alpha_\min}{R_1} \bigg)^2 
  + \bigg( \myfrac{m - \beta_\min}{R_2} \bigg)^2
\qquad  \hbox{where ~} -\infty < n,m < + \infty ~.
\label{spectrumHiggs1}
\eeqn
There are four zero modes associated with $A_{y^I}^a$
for $(\alpha_\min, \beta_\min) = (0,0)$
$(mod ~ 1)$, while only two otherwise.  These zero modes become massive 
at the quantum level.

\vskip .2cm
\noindent
{\bf Case 1.}  $(\alpha_\min, \beta_\min) = (0,0)$ $(mod ~ 2)$

In this case there remains $U(1)$ symmetry.   There are four zero modes
associated with $A_{y^1}^1, A_{y^2}^1, A_{y^1}^2, A_{y^2}^2$.  The effective
potential is given by
\beqn
&&\hskip -1cm
\hV_\eff [A^1_{y^1}, A^1_{y^2}, A^2_{y^1}, A^2_{y^2}]
= \hV_\eff^{\rm 1-loop}   \cr
\noalign{\kern 10pt}
&&\hskip 0.cm
+ g^2 \Big\{ (A^1_{y^1})^2 (A^2_{y^2})^2 + (A^2_{y^1})^2 (A^1_{y^2})^2
- 2 A^1_{y^1} A^2_{y^1} A^1_{y^2} A^2_{y^2}  \Big\} ~~,
\label{effectiveV1}
\eeqn
where
the second term comes from $\onehalf \Tr (F_{y^1 y^2})^2$
at the tree level.  
The evaluation of $\hV_\eff^{\rm 1-loop}$ for general configurations
with $F_{y^1 y^2} \not= 0$ is difficult.  We observe that the mass spectrum is
$U(1)$ symmetric and expect that fluctuations with vanishing $F_{y^1 y^2}$
form a normal basis for the zero modes.  We therefore make an approximation
\beeq
\hV_\eff^{\rm 1-loop} \sim
V_\eff[\alpha , \beta]
\label{effectiveV2}
\eneq
where  $V_\eff[ \alpha ,  \beta]$ is the effective potential obtained
in the preceding sections with
$\alpha = 2gR_1 \sqrt{  (A^1_{y^1})^2 +  (A^2_{y^1})^2 }$ and
$\beta = 2gR_2 \sqrt{  (A^1_{y^2})^2 +  (A^2_{y^2})^2 }$.

As an example, take the pure gauge theory.  The effective potential
is given by $V_\eff[\alpha , \beta] = 4 I(\alpha,\beta)$.  (See
(\ref{potentail-gauge3}).)  The mass matrix is given by the second derivatives
of $\hV_\eff$ with respect to $A_{y^I}^a$ evaluated at vanishing $A_{y^I}^a$.
One finds that 
\beeq
(\hbox{mass})^2 = \cases{
8 \pi^2 g_4^2  R_1^3 R_2 \myfrac{\dd^2 V_\eff}{\dd \alpha^2} 
  \Bigg|_{(\alpha,\beta)=(0,0)}  &for $A^1_{y^1}$, $A^2_{y^1}$ ~,\cr
\noalign{\kern 10pt}
8 \pi^2 g_4^2  R_1 R_2^3 \myfrac{\dd^2 V_\eff}{\dd \beta^2} 
  \Bigg|_{(\alpha,\beta)=(0,0)}  &for $A^1_{y^2}$, $A^2_{y^2}$ ~.\cr}
\label{spectrumHiggs2}
\eneq
Here the four-dimensional gauge coupling is given by
$g_4^2 = g^2/2 \pi^2 R_1 R_2$.    We used the fact 
$\dd^2 V_\eff / \dd \alpha \dd \beta |_{(\alpha,\beta)=(0,0)} = 0$.
When $R_1=R_2$, the masses are given by 
\beeq
(\hbox{mass})^2 = 
\myfrac{8  C_1 g_4^2}{\pi^5 R^2}  ~~,~~
C_1 = \sum_{n=1}^\infty \myfrac{1}{n^4} + 
 \sum_{n=1}^\infty \sum_{m=1}^\infty \myfrac{1}{(n^2+m^2)^2} 
\approx 1.507 
\label{spectrumHiggs3} 
\eneq
for all zero modes.

\vskip .2cm
\noindent
{\bf Case 2.}  $(\alpha_\min, \beta_\min) = (1,1)$ $(mod ~ 2)$

In the example discussed in the subsection 5.2, the global minimum of the 
$V_\eff(\alpha,\beta)$ is located at $(\alpha_\min, \beta_\min) = (1,1)$.
In the new gauge $(P_0, P_1, P_2) = (\tau^3, -\tau^3, -\tau^3)$.
There are no zero modes for the fermions in the fundamental representation
with $(\delta_1, \delta_2)=(0,0)$.  

There still remains the $U(1)$ symmetry.  The masses of the four zero modes
associated with $A_{y^I}^a$ are given by (\ref{spectrumHiggs2}) with
$V_\eff$ in (\ref{potential-fF1}).  For $R_1=R_2=R$ they are given by
\beqn
&&\hskip -1cm
(\hbox{mass})^2 = 
\myfrac{2 (4 C_1 + N_{00}^{f,F} C_2)g_4^2}{\pi^5 R^2}  ~~,~~ \cr
\noalign{\kern 5pt}
&&\hskip -1cm
C_2 = \sum_{n=1}^\infty \myfrac{(-1)^{n-1}}{n^4} 
 - \sum_{n=1}^\infty \sum_{m=1}^\infty 
   \myfrac{(-1)^{n+m}}{(n^2+m^2)^2} 
\approx 0.753  ~~.
\label{spectrumHiggs4} 
\eeqn

\vskip .2cm
\noindent
{\bf Case 3.}  $(\alpha_\min, \beta_\min) \not= (0,0)$ $(mod ~ 1)$

The examples discussed in the subsections 5.3 and 5.4 belong to this 
category.  There are only two zero modes associated with $A_{y^1}^1$ and 
$A_{y^2}^1$.  The lightest modes of $Z_2$ doublet 
$(A_{y^I}^2, A_{y^I}^3)$ has 
(mass)$^2 = (\bar \alpha_\min/R_1)^2 + (\bar \beta_\min/R_2)^2$ where
$\bar \alpha_\min$ and $\bar \beta_\min$ are the distances to the nearest 
integers of $\alpha_\min$ and $\beta_\min$, respectively. 

The masses of the two zero modes of $A_{y^I}^1$ are evaluated from
$\hV_\eff(A_{y^1}^1,A_{y^2}^1) = V_\eff(\alpha_\min + 2gR_1 A_{y^1}^1,
\beta_\min + 2gR_2 A_{y^2}^1)$.  Take the example in the subsection 5.4
with $N_{00}^{f,F}=0$ and $N_{01}^{f,Ad} \ge 2$.
The global minimum is located at $(\alpha_\min,\beta_\min)=(\onehalf, 0) ~~
(mod ~ 1)$. It follows from (\ref{potential-fFAd1}) that, for $R_1=R_2=R$,
\beqn
&&\hskip -1cm
(\hbox{mass})^2 = \cases{
(- C_3 + N_{01}^{f,Ad} C_2) \myfrac{8 g_4^2}{\pi^5 R^2} &for $A_{y^1}^1$~,\cr
(+ C_4 + N_{01}^{f,Ad} C_2) \myfrac{8 g_4^2}{\pi^5 R^2} &for $A_{y^2}^1$~,\cr}
 \cr
\noalign{\kern 5pt}
&&\hskip -1cm
C_3 = \sum_{n=1}^\infty \myfrac{(-1)^{n-1}}{n^4} 
 + 2 \sum_{n=1}^\infty \sum_{m=1}^\infty 
   \myfrac{(-1)^{n-1} n^2}{(n^2+m^2)^3} 
\approx 1.152  ~~, \cr
\noalign{\kern 5pt}
&&\hskip -1cm
C_4 = \sum_{m=1}^\infty \myfrac{1}{m^4} 
 +2 \sum_{n=1}^\infty \sum_{m=1}^\infty 
   \myfrac{(-1)^n m^2}{(n^2+m^2)^3} 
\approx 0.776  ~~.
\label{spectrumHiggs5} 
\eeqn

\vskip .5cm
\leftline{\bf 6.2 Four-dimensional fermions}

From the
phenomenological viewpoint it is necessary to accommodate fermions
with small, but nonvanishing masses.  In the four-dimensional
standard model of electroweak interactions, Yukawa interactions
provide such small masses.  In higher dimensional gauge theory, however,
Yukawa interactions are sometimes absent, or a part of gauge interactions
so that it becomes difficult to allow small, but nonvanishing fermion
masses.

We would like to point out that such small masses might be accommodated 
in the framework of gauge theory on orbifolds through the combination 
of $T^2$ twists and dynamics of Wilson line phases.  At the moment
such scenario is realized only if special combinations of matter
fields are arranged.   It might occur naturally in supersymmetric
theories.  We reserve discussions of supersymmetric theories for
the future publication.

The models discussed in the subsection 5.5 give  nice examples.
In the model described by (\ref{potential-fFAd2}) with 
$(N^{f,Ad}_{00}, N^{f,Ad}_{01}, N^{f,F}_{00, \rm doublet}) =(2, 0, 3)$, 
 one of the global minima of the effective potential is located
at $(\alpha,\beta)=(0.5, 0.5)$ and $(0.486,0.486)$
for $(a,b)=(0.5, -0.5)$ and $(0.51, -0.51)$, respectively. Fermions in the
fundamental representation have the mass spectrum given by 
(\ref{spectrum-doublet1}) with $\delta_j=0$.  The relevant parameters are
$(a + \alpha, b + \beta)$ and $(a - \alpha, b - \beta)$.  Unless
one of these two pairs has elements equal or close to even integers,
fermions acquire masses of $O(R_1^{-1})$ or $O(R_2^{-1})$.  We see that
none of four-dimensional fermions in the model are light.
In the model with 
$(N^{f,Ad}_{00}, N^{f,Ad}_{01}, N^{f,F}_{00, \rm doublet}) =(0, 3, 3)$,
the situation does not change.   There are no light fermions in four dimensions.
For $(a,b) = (0.5, 0)$, the global minima of $V_\eff$ are located at
$(\alpha, \beta) = (\pm 0.5, 1)$, whereas for $(a,b) = (0.5, 1)$
they are located at $(\alpha, \beta) = (\pm 0.5, 0)$.

This is a general feature.  
Fermions either in $Z_2$ singlets or in $Z_2$ doublets give contributions
to the effective potential for Wilson line phases such that the effective
potential is minimized by four-dimensional massive fermions, as can 
be inferred from (\ref{potential-fFAd2}).  The tendency is reversed 
by contributions from bosons.  In supersymmetric theories contributions 
from bosons and fermions cancel if  supersymmetry remains unbroken.
When supersymmetry is softly broken as in the Scherk-Schwarz
breaking, nontrivial dependence of the effective potential on
twist parameters and Wilson line phases appears.\cite{Takenaga1, HHHK}
  Then fermions in four
dimensions may have small nonvanishing masses.

\sxn{Conclusions and discussion}
We have studied gauge theory with matter on $M^4\times T^2/Z_2$. 
We have classified general boundary conditions for fields 
on the orbifold $T^2/Z_2$. The equivalence relation among various sets of
boundary conditions holds as a result of the existence of 
boundary-condition-changing gauge transformations.  By incorporating 
Wilson line degrees of freedom correctly, one can establish 
the same physics in each 
equivalence class of boundary conditions.

The $Z_2$-orbifolding boundary conditions, which are specified by  
parity matrices $P_i$ $(i=0, 1, 2)$, break the gauge symmetry at the tree 
level. In order to find physical symmetry of the
theory at low energies, which, in general, is different from the symmetry 
of  boundary conditions, one must take into account dynamics of Wilson
line phases by the  Hosotani 
mechanism, through which further gauge symmetry breaking can be induced
at the quantum  level. 

We have  studied the $SU(2)$ gauge theory in detail to clarify
physical symmetry  at low energies.  We have chosen  
boundary conditions  of the $Z_2$ orbifolding that break the $SU(2)$
gauge symmetry  down to $U(1)$. Depending on the matter content, 
the residual $U(1)$ gauge symmetry is further broken through the
Hosotani mechanism and the original
$SU(2)$ gauge symmetry is completely broken. This indicates that 
the electroweak gauge symmetry 
breaking $SU(2)_L\times U(1)_Y \rightarrow U(1)_{em}$ 
can be realized by the Hosotani mechanism, once a larger gauge group 
is chosen to start with.  Indeed, such implementation of symmetry 
breaking has been attempted in the literature 
under the  name of the gauge-Higgs unification. 
The $SU(6)$ model on $M^4 \times  (S^1/Z_2)$ realizes such a
scenario.\cite{gaugeHiggs3}

Regarding gauge symmetry breaking, the study in the
present paper has been limited mostly to  the 
case where the ratio of the
size of the two extra dimensions are equal $r\equiv R_2/R_1=1$. 
Varying $r$ modifies the shape of 
the effective potential to give different gauge symmetry breaking 
patterns. This study may be important in the model building.  
One can introduce two distinct scales, the GUT scale and electroweak scale.

We have also studied the particle spectrum in four dimensions.
Some of the extra-dimensional components of gauge fields,
four-dimensional `Higgs' scalar fields, are massless at the tree level, 
but become massive by radiative corrections.  Their typical mass is
given  by $g_4/R_1$ or $g_4/R_2$, where $g_4$ is the four-dimensional gauge
coupling constant. 

It is interesting to extend our work to  higher rank gauge 
groups and to study more realistic models of  gauge symmetry 
breaking  and  gauge-Higgs unification. 
It is particularly important to consider supersymmetric gauge
theory in this framework.  A realistic fermion mass spectrum in four
dimensions might be achieved in supersymmetric theories as a result of
dynamics of Wilson line phases, additional $T^2$ twists on matter fields,
and supersymmetry breaking.  We hope to come back to this point in the near
future.

\vskip 1cm

\leftline{\bf Acknowledgments}
This work was supported in part by  Scientific Grants from the Ministry of 
Education and Science, Grant No.\ 13135215, Grant No.\ 13640284, and
Grant No.\ 15340078 (Y.H), and by the 21st Century COE Program at Osaka 
University (K.T.).


\vskip 1.cm

\def\jnl#1#2#3#4{{#1}{\bf #2} (#4) #3}

\def\Zphys{{\em Z.\ Phys.} }
\def\jssc{{\em J.\ Solid State Chem.\ }}
\def\jpsJ{{\em J.\ Phys.\ Soc.\ Japan }}
\def\ptps{{\em Prog.\ Theoret.\ Phys.\ Suppl.\ }}
\def\PTP{{\em Prog.\ Theoret.\ Phys.\  }}

\def\JMP{{\em J. Math.\ Phys.} }
\def\NPB{{\em Nucl.\ Phys.} B}
\def\NP{{\em Nucl.\ Phys.} }
\def\PLB{{\em Phys.\ Lett.} B}
\def\PL{{\em Phys.\ Lett.} }
\def\PRL{\em Phys.\ Rev.\ Lett. }
\def\PRB{{\em Phys.\ Rev.} B}
\def\PRD{{\em Phys.\ Rev.} D}
\def\PRe{{\em Phys.\ Rep.} }
\def\AP{{\em Ann.\ Phys.\ (N.Y.)} }
\def\RMP{{\em Rev.\ Mod.\ Phys.} }
\def\ZPC{{\em Z.\ Phys.} C}
\def\SCI{\em Science}
\def\CMP{\em Comm.\ Math.\ Phys. }
\def\MPLA{{\em Mod.\ Phys.\ Lett.} A}
\def\IJMPA{{\em Int.\ J.\ Mod.\ Phys.} A}
\def\IJMPB{{\em Int.\ J.\ Mod.\ Phys.} B}
\def\EPJC{{\em Eur.\ Phys.\ J.} C}
\def\PR{{\em Phys.\ Rev.} }
\def\JHEP{{\em JHEP} }
\def\cmp{{\em Com.\ Math.\ Phys.}}
\def\JPA{{\em J.\  Phys.} A}
\def\JPG{{\em J.\  Phys.} G}
\def\NJP{{\em New.\ J.\  Phys.} }
\def\CQG{\em Class.\ Quant.\ Grav. }
\def\ATMP{{\em Adv.\ Theoret.\ Math.\ Phys.} }
\def\ibid{{\em ibid.} }

\renewenvironment{thebibliography}[1]
         {\begin{list}{[$\,$\arabic{enumi}$\,$]}  
         {\usecounter{enumi}\setlength{\parsep}{0pt}
          \setlength{\itemsep}{0pt}  \renewcommand{\baselinestretch}{1.2}
          \settowidth
         {\labelwidth}{#1 ~ ~}\sloppy}}{\end{list}}

\end{document}